\documentclass{ws-p8-50x6-00}

\begin{document}

\title{The Theory of Atom Lasers}

\author{R. Ballagh}

\address{Physics Department, University of Otago, Dunedin, New Zealand
\\E-mail: ballagh@physics.otago.ac.nz
\\Web: www.physics.otago.ac.nz/research/bec
}

\author{C.M. Savage}

\address{Physics Department, The Australian National University,
\\ Canberra, ACT 0200, Australia.
\\ E-mail: craig.savage@anu.edu.au
\\ Web: www.anu.edu.au/physics/Savage}  


\maketitle

\abstracts{We review the current theory of atom lasers. A tutorial
treatment of second quantisation and the  Gross-Pitaevskii equation is
presented, and  basic concepts of coherence are outlined. The generic
types of atom laser models are surveyed and illustrated by specific
examples. We conclude with detailed treatments of the mechanisms of gain
and output coupling.}

\section{Introduction}

\label{sec.Introduction}

An atom laser is a device which produces a bright, directed, and coherent
beam of atoms. An ideal atom laser beam is a single frequency de Broglie
matter wave, approximating a noiseless sinusoidal wave; in particular it
will have a well defined frequency, phase, and amplitude. Quantum
mechanically, such a wave is a mode of the quantised field, and the
properties of brightness and well defined phase require the mode to be
highly occupied. This is equivalent to requiring Bose degeneracy, and rules
out the possibility of a fermionic atom laser\cite{Wiseman97}.

Atom lasers are a recent concept, and to our knowledge the earliest
scientific description appeared in 1993 \cite{Savage93}. The first
comprehensive papers on matter wave amplification\cite{Borde} and atom lasers%
\cite{Olshanii95,Spreeuw95,Holland96} appeared in 1995, and the first
experimental demonstration of an atom laser in 1997 \cite{Mewes97}. The
essential components of an atom laser have been identified by analogy with
optical lasers and are illustrated in Fig.~1. 
\begin{figure}[tbp]
\centerline{\epsfig{file=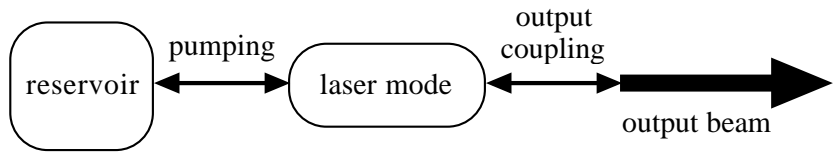,width=4.4in}}
\caption{Schematic diagram of a generic atom laser.}
\end{figure}
They are; a reservoir of atoms from which the laser is pumped, a laser mode
in which atoms are accumulated, a pumping (or gain) mechanism which
transfers atoms from the reservoir into the laser mode by stimulated
transitions, and an output coupler which produces an output beam from the
laser mode while preserving its coherence.

Many of the possible applications of atom lasers are probably still to be
imagined, nevertheless atom lasers have the potential to revolutionise atom
optics just as optical lasers did conventional optics\cite%
{Helmerson,Ketterle}. A bright coherent beam greatly enhances
interferometric capabilities, for example allowing fringes to be obtained on
a small time scale, and the use of unequal path length interferometers. The
wavelength of the atoms is typically much smaller than light, and an obvious
potential technological application is the precision deposition of matter
for nanofabrication. While the similarities to optical lasers are important,
the fundamental differences between atoms and photons are also significant.
Atoms interact with each other, are subject to gravity, have complex
internal structure, and may travel at any speed, not just $c.$ Their
interactions (collisions) make atom optics intrinsically nonlinear, as 
illustrated by a recent experiment showing self-induced four-wave mixing of
atom laser beams\cite{Deng99}. Their interaction with the gravitational
field gives the opportunity for atom laser beams to be used in high
precision gravitational and inertial measurements\cite{Chu}, and very
sensitive atom interferometers may be able to detect changes in space-time%
\cite{Helmerson}. Eventually we expect that atom lasers will provide a
convenient source for greatly improved atomic clocks, or for manipulation of
the quantum mechanical state of the matter wave, analogous for example to
optical squeezing\cite{Bachor}.

As of April 2000 five experimental groups have demonstrated atom lasers
experimentally, in either pulsed operation\cite{Mewes97,Anderson98,Martin98}
or with quasi-continuous output\cite{Bloch99,Hagley99}. In each case, the
laser mode consists of a Bose-Einstein condensate confined in the ground
state of a magnetic trap, gain is achieved by evaporatively cooling a
reservoir of thermal atoms, and the output coupling is accomplished by using
an external electromagnetic field to transfer the internal state of the atom
from a trapped to an untrapped state. In these experiments the output
coupling and the pumping processes are separated in time. Q-switching in
pulsed optical lasers also separates the pumping and output coupling, but
the cycle time is much shorter. Hence, a closer analogy is to a cavity
loaded with coherent light which then leaks out through a mirror. The use of
the name ``atom laser'' is justified by the coherence of the output beam,
which was first demonstrated by the MIT group\cite{Mewes97,KetterleMiesner},
and is due fundamentally to the stimulated emission of bosonic atoms into
the ground state of the trap. It remains an outstanding experimental goal to
achieve simultaneous pumping and output coupling to make a truly continuous
atom laser.

Despite the parallels with optical laser theory, the fundamental
differences between photons and atoms remain important. They arise
principally because atoms have rest mass, and because they interact with
each other. Unlike photons, atoms cannot be created or destroyed, rather
they are transferred into the laser mode from some other mode. The
interactions produce phase dynamics, which degrades the coherence of the
atom laser, and self repulsion which spreads the output beam and limits
the possible focussing. Even when these interactions are neglected,
atoms have dispersive propagation  in vacuum.

Various critical parameters for atom and optical lasers are radically
different. For example the characteristic frequency in the optical case is
the frequency of the light, of order $10^{15}$ Hz, while for atoms the de
Broglie frequency is closer to that of the trapping potential, of order $%
10^{3}$ Hz. And, whereas it is so far only practical to get atoms to
accumulate into the ground mode of a trap, optical lasers typically operate
on high modes of the optical cavity. Finally, a practical consideration is
that  atom lasers can operate only in ultra high vacuum.

In this paper we review the current state of theory of atom lasers. We
begin, in sections 2-4, with a brief tutorial review of second quantisation
and the Gross-Pitaevskii equation, and the concepts of coherence.  In
section 5 we survey the main types of theoretical models for atom lasers,
and illustrate these with specific examples. Finally, in sections 6 and 7 
we provide more detailed treatments of the mechanisms of gain and output
coupling. 

%%%%%%%%%%%%%%%%%%%%%%%%%%%%%%%%%%%%%%%%%%%%%%%%%%%%%%%%%%%%%%%%%%%%%%%

\section{Second Quantization}

In this section we review the theory of quantum fields: the most fundamental
theoretical framework for treating atom lasers. As long as their internal
atomic structure is not disrupted, atoms may be approximated as
field quanta. However the internal state of the atom may
change, e.g. between trapped and untrapped orientational spin states, and
then each relevant internal state is described by its own field.

Identity of particles arises naturally in quantum field theory, as all
particles of the same type are indistinguishable quantised excitations of
the same field. We introduce quantum fields by second quantization of the
wavefunction. For a more complete treatment see {\it Quantum Mechanics} 
by E.~Merzbacher.
The Hamiltonian for $N$ particles interacting
via the pairwise potential $V\left( \mathbf{r},\mathbf{r}^{\prime }\right)$
is 
\begin{equation}
H =\sum_{i=1}^{N}H_{0}\left( \mathbf{r}_{i}\right) + \frac{1}{2}
\sum_{i,j} V\left( \mathbf{r}_{i},\mathbf{r}_{j} \right) ,
\label{eq:hamiltonian}
\end{equation}
where the $\mathbf{r}_{i}$ are the particle position vectors. The factor of
a half in the 2-body interaction term corrects for double counting of
particle pairs. The single particle Hamiltonian $H_{0}\left( \mathbf{r}%
\right) $ typically represents the kinetic energy and an external potential $%
V_{E}\left(\mathbf{r}\right) $, such as a trap. In the position
representation it is 
\begin{equation}
H_{0}=-\frac{\hbar^{2}}{2m}\bigtriangledown ^{2}+V_{E}\left( \mathbf{r}
\right) .  \label{eq:single particle H}
\end{equation}
Second quantization allows us to automatically account for the
identity of the particles, and is achieved by introducing the operator
field $\Psi \left( \mathbf{r}\right) $ 
which annihilates a particle at position $\mathbf{\ r}$. 
It obeys the following canonical commutation rules with its hermitian
conjugate particle creation field $\Psi ^{\dagger }\left( \mathbf{r}\right)$, 
\begin{equation}
\left[ \Psi \left( \mathbf{r}\right) ,\Psi ^{\dagger }\left( \mathbf{r}
^{\prime }\right) \right] =\delta \left( \mathbf{r}-\mathbf{r} ^{\prime
}\right) ,\qquad \left[ \Psi \left( \mathbf{r}\right) ,\Psi \left( \mathbf{r}%
^{\prime }\right) \right] =0 .  \label{eq:commutators}
\end{equation}
The second quantized Hamiltonian has the form of an ``expectation value'' of
the Hamiltonian (\ref{eq:hamiltonian}) with respect to these fields; 
\begin{equation}
H=\int \Psi ^{\dagger }\left( \mathbf{r}\right) H_{0}\Psi \left( \mathbf{r}
\right) d^{3}r 
+\frac{1}{2} \int \int \Psi ^{\dagger }\left( 
\mathbf{r}\right) \Psi ^{\dagger }\left( \mathbf{r}^{\prime }\right) V\left( 
\mathbf{r}, \mathbf{r}^{\prime }\right) \Psi \left( \mathbf{r}^{\prime
}\right) \Psi \left( \mathbf{r}\right) d^{3}{r}^{\prime }d^{3}r.  
\label{eq:2nd quantised H}
\end{equation}
In the Heisenberg picture, operators obey the usual Heisenberg equation of
motion. For example the field operator obeys 
\begin{equation}
i\hbar \frac{d\Psi \left( \mathbf{r},t\right) }{dt}=\left[ \Psi \left( 
\mathbf{r},t\right) ,H\right] .  \label{eq:heisenberg}
\end{equation}
Using the commutation rules (\ref{eq:commutators}) and the Hamiltonian (\ref%
{eq:2nd quantised H}) this becomes 
\begin{equation}
i\hbar \frac{d\Psi \left( \mathbf{r}\right) }{dt}=\left\{ H_{0}\left( 
\mathbf{r}\right) +\int \Psi ^{\dagger }\left( \mathbf{r}^{\prime }\right)
V\left( \mathbf{r},\mathbf{r}^{\prime }\right) \Psi \left( \mathbf{r}%
^{\prime }\right) 
d^{3}{r}^{\prime }\right\} \Psi \left( \mathbf{r}%
\right) .  \label{eq:field dynamics}
\end{equation}
It is often useful to expand the field operators in terms of products of
operators $a_{i}$ and the elements of a basis set of orthonormal mode
functions $\left\{ \phi _{i}\left( \mathbf{r}\right) \right\}$, 
\begin{equation}
\Psi \left( \mathbf{r}\right) =\sum_{i}a_{i}\phi _{i}\left( \mathbf{r}
\right) \Longrightarrow a_{i}=\int \phi _{i}^{*}\left( \mathbf{r}\right)
\Psi \left( \mathbf{r}\right) d^{3}{r}  \label{eq:modes}
\end{equation}
This expansion is the basis of the Fock, or particle number, representation
of the field. The commutation relations for the annihilation and creation
operators follow from Eqs.(\ref{eq:commutators}) and the mode function
orthonormality, 
\begin{equation}
\left[ a_{i},a_{j}^{\dagger }\right] =\int \int \phi _{i}\left( \mathbf{r}%
\right) \phi _{j}^{*}\left( \mathbf{r}^{\prime }\right) \left[ \Psi \left( 
\mathbf{r}\right) ,\Psi ^{\dagger }\left( \mathbf{r}^{\prime }\right) \right]
d^{3}{r}d^{3}{r}^{\prime } = \delta_{ij}
\label{eq:a commutator}
\end{equation}
These commutators are those for a set of harmonic oscillators. Hence we can
identify $a_{i}$ as a harmonic oscillator type annihilation operator for the
mode $\phi_{i}$, and its Hermitian conjugate $a^{\dagger}_{i}$ as the
corresponding creation operator. In quantum field theory these are
interpreted as annihilating and creating field quanta, which in our case are
atoms, in the mode. The quantum mechanical state of the system is that of
these oscillators. In the Fock representation the second quantised
Hamiltonian (\ref{eq:2nd quantised H}) becomes 
\begin{eqnarray}
&&H = \sum_{i,j}a_{i}^{\dagger }a_{j}\int \phi _{i}^{*}\left( \mathbf{r}
\right) H_{0}\left( \mathbf{r}\right) \phi _{j}\left( \mathbf{r}\right) d^{3}%
{r} \\
&&+\frac{1}{2} \sum_{i,j,k,l}a_{i}^{\dagger }a_{j}^{\dagger }a_{k}a_{l}\int
\int \phi _{i}^{*}\left( \mathbf{r}\right) \phi _{j}^{*}\left( \mathbf{r}%
^{\prime} \right) V\left( \mathbf{r}, \mathbf{r}^{\prime }\right) \phi
_{k}\left( \mathbf{r} ^{\prime }\right) \phi _{l}\left( \mathbf{r}\right)
d^{3}{r}d^{3}{r}^{\prime }  \nonumber  \label{eq:H using as}
\end{eqnarray}
If the mode functions are chosen to be eigenstates of the single particle
free Hamiltonian, $H_{0}\phi _{i}=E_{i}\phi _{i}$, the free part of the
Hamiltonian can be expressed in terms of the oscillator number operators $%
\{a_{i}^{\dagger } a_{i}\}$. Denoting the integrals in the interaction part
of the Hamiltonian by $V_{ijkl}$ we then find 
\begin{equation}
H=\sum_{i}a_{i}^{\dagger }a_{i}E_{i} +\frac{1}{2} \sum_{i,j,k,l}a_{i}^{%
\dagger }a_{j}^{\dagger }a_{k}a_{l}V_{ijkl}.  \label{eq:H free basis}
\end{equation}
Instead of solving for the full dynamics one may find the quasi-particles:
the weakly interacting excitations of the system. 
The aim is to find an (approximately) diagonal form of  $H$. 
The field is expressed in
terms of the quasi-particle creation $b_{i}^{\dagger }$ and annihilation $%
b_{i}$ operators by 
\begin{equation}
\Psi \left( \mathbf{r}\right) =\sum_{i}\left( u_{i}\left( \mathbf{r}%
\right) b_{i}-v_{i}^{*}\left( \mathbf{r}\right) b_{i}^{\dagger }\right) +\psi
_{C}\left( \mathbf{r}\right) .  \label{eq:quasiparticles}
\end{equation}
$\psi _{C}$ is called the condensate, or macroscopic, wavefunction. It is
normalised so that the integral over all space of its squared modulus is $%
N_{C}$, the number of particles in the condensate. Assuming a non-zero
condensate wavefunction is the basis of the ``spontaneous symmetry
breaking'' method\cite{Huang87} for analysing BECs, which is discussed
further in the next section. The functions $u_{i}\left( \mathbf{r}\right) $
and $v_{i}\left( \mathbf{r} \right)$ are determined such that the
Hamiltonian (\ref{eq:H free basis}) has the approximate and noninteracting
form 
\begin{equation}
H\approx \sum_{i}b_{i}^{\dagger }b_{i}E_{i}^{\prime } 
+ \mbox{\rm constant}
\label{eq:quasiparticle H}
\end{equation}
The quanta of these modes are the quasi-particles. The transformation to a
quasi-particle representation is known as the Bogoliubov transformation.

\section{The Gross-Pitaevskii Equation}

\label{sec:Interactions}

In an atom laser the atoms will interact, both in the laser itself and in
the output atomic beam. The interactions between cold atoms are usefully
described as collisions. Elastic collisions conserve the total external
energy of the colliding atoms. Inelastic collisions do not, and energy may
be exchanged between the external and internal degrees of freedom. Changes
in internal state due to inelastic collisions are an important source of
losses of atoms from magnetic traps. We consider a simple s-wave scattering
model of elastic collisions. This approximation is valid for low
temperatures and low densities, such that both the atomic de Broglie
wavelength and the inter-particle separation exceeds the characteristic
range of the interaction potential. For further discussion see the 
article by Burnett in this volume.

In this limit the atom-atom potential may be approximated by a delta
function potential with the same interaction energy 
\begin{equation}
V\left( \mathbf{r},\mathbf{r}^{\prime }\right) \approx U_{0}\delta \left( 
\mathbf{r}-\mathbf{r}^{\prime }\right) ,\qquad U_{0}=\frac{4\pi \hbar^{2} a}{%
m}.  \label{eq:delta fn potential}
\end{equation}
$a$ is called the ``scattering length''. It is positive for repulsive
interactions and negative for attractive ones. For indistinguishable
particles the scattering cross section (the ratio of scattered to incident
particle flux) is $\sigma =8\pi a^{2}$, rather than $4\pi a^{2}$, as found
for distinguishable particles. $U_{0}$ is the atom-atom interaction energy
per atom. Using this approximation in the field operator equation (\ref%
{eq:field dynamics}) gives 
\begin{equation}
i\hbar \frac{d\Psi \left( \mathbf{r}\right) }{dt}=\left\{ H_{0}\left( 
\mathbf{r}\right) +U_{0}\Psi ^{\dagger }\left( \mathbf{r}\right) \Psi \left( 
\mathbf{r}\right) \right\} \Psi \left( \mathbf{r}\right) .
\label{eq:heisenbergfield}
\end{equation}
We next approximate this operator equation by a non-linear Schrodinger
equation for the condensate wavefunction.

The condensate wavefunction $\psi _{C}$ is defined by Eq.(\ref%
{eq:quasiparticles}) to be the contribution to the expectation value of the
field operator $\Psi $ due to the quasi-particle vacuum state. According to
the ``spontaneous symmetry breaking'' assumption, above the BEC transition
temperature $\psi _{C}=0$, while below this temperature it is nonzero\cite%
{Huang87}. The name arises because the Hamiltonian (\ref{eq:2nd quantised H}%
) is symmetric, or invariant, under a global phase change $\Psi \left( 
\mathbf{r}\right) \rightarrow $ $\Psi \left( \mathbf{r} \right) e^{i\theta }$%
, whereas the condensate wavefunction must have a particular phase, breaking
the global phase symmetry. This symmetry is a consequence of the pairing of
particle creation and annihilation operators, and hence of particle number
conservation, in the Hamiltonian (\ref{eq:2nd quantised H}). 
Particle number conservation in fact implies that $\Psi(r)=0$
always\cite{Molmer,Gardiner}, but spontaneous
symmetry breaking is a useful heuristic assumption, which used with due
caution, gives experimentally verified results. 
In fact, a number of calculations have been made without this assumption, 
and obtained the same results. Griffin's article in this volume
further explores the spontaneous symmetry breaking assumption.

At zero temperature there are no quasi-particle excitations and the field
operator $\Psi$ in Eq.(\ref{eq:heisenbergfield}) may be approximated by the
single particle condensate wavefunction $\psi_{C}$. We further assume that products of field
operators may be approximated by the corresponding products of condensate
wavefunctions.  The condensate wavefunction then obeys the
Gross-Pitaevskii equation\cite{Lifshitz80} 
\begin{equation}
i\hbar \frac{\partial \psi _{C}\left( \mathbf{r}\right) }{\partial t}=
\left\{ -\frac{\hbar
^{2}}{2m}\nabla ^{2}+V_{E}\left( \mathbf{r}\right) +U_{0}\left| \psi
_{C}\left( \mathbf{r}\right) \right| ^{2}\right\} \psi _{C}\left( \mathbf{r}
\right) .  \label{eq:GP eqn}
\end{equation}
Mathematically, this equation is a non-linear Schr\"{o}dinger equation
about which much is known.  Nevertheless, in general a numerical
solution is required\cite{Edwards95}.

\section{Coherence}

\label{sec:coherence}

An atom laser is a device which emits a coherent beam of atoms. Coherence
and noise are closely related. A noiseless classical wave, that is a perfect
single frequency sinusoid, is defined to be perfectly coherent or perfectly
correlated. For such a wave the field amplitude at any space-time event $%
A(x,t)$ completely determines the amplitude at any other event. In practice
the field amplitude fluctuates due to classical or quantum mechanical noise,
and such determination is not possible. Classical noise may be technical,
such as vibrating components, or dynamical, such as relaxation oscillations
in optical lasers\cite{Siegman}. Quantum mechanical noise may be interpreted
as a consequence of the uncertainty relations.

Optical lasers operating far above threshold have a well stabilised
intensity. Semiclassical theories describe optical lasers by quantising the
atoms only, not the light. Introducing a phenomenological noise term allows
spontaneous emission to be modelled without quantising the light. For such a
theory, using the dimensionless intensity $I$ defined by Mandel and Wolf\cite%
{Mandel&Wolf}, far above threshold 
\begin{equation}
\langle ( \Delta I )^{2} \rangle \approx 2 , \quad \langle ( \Delta I )^{2}
\rangle^{1/2} / \langle I \rangle \approx (\sqrt 2) / \langle I \rangle
\label{Eq.laserintensity}
\end{equation}
The angle brackets indicate the average over the noise which models
spontaneous emission. The inverse dependence of the relative intensity
fluctuations on the intensity is characteristic of the optical laser. It is
expected to be true for the atom flux from atom lasers. In contrast, for
thermal beams the relative intensity fluctuations are constant.

The phase of an optical laser far above threshold is also stabilised. This
is measured by the electric field amplitude temporal coherence function $%
\langle E^{*}(t_{1}) E(t_{2}) \rangle$. Its Fourier transform is the
spectral density, which is approximately Lorentzian\cite{Mandel&Wolf}. Its
width is the laser bandwidth, which is found to be inversely proportional to
the intensity\cite{Schawlow&Townes}. This is another characteristic optical
laser property which we would expect to be true for atom lasers.

We now describe the coherence functions which are used to quantify the
coherence of fields. The amplitude coherence function is: 
\begin{equation}
\langle A^{*}(\mathbf{x}_{1}) A(\mathbf{x}_{2}) \rangle ,  \label{Eq.clcoh}
\end{equation}
where the angle brackets denote an average over noise, and a bold $\mathbf{x}%
_{k}$ denotes the spacetime event $\mathbf{x}_{k}=(x_{k},t_{k})$. This
coherence function is unambiguously referred to as a two-time amplitude
correlation function. In the literature it may be referred to as either a
first\cite{WallsMilb} or second order\cite{Mandel&Wolf} amplitude coherence
function. We use the former convention. If this function factorises 
\begin{equation}
\langle A^{*}(\mathbf{x}_{1}) A(\mathbf{x}_{2}) \rangle = \langle A^{*}(%
\mathbf{x}_{1}) \rangle \: \langle A(\mathbf{x}_{2}) \rangle
\label{Eq.clcohfact}
\end{equation}
the amplitude is said to be first order amplitude coherent. This definition
captures the concept that a coherent field has no more noise in amplitude
products than that due to the noise in the amplitude itself.

The definition of coherence as the degree to which coherence functions
factorise is retained for quantum fields. Coherence functions (also called
correlation functions) of order $n$ were defined by Glauber\cite{Glauber} in
terms of the field operators by 
\begin{equation}
G^{(n)}(\mathbf{x}_{1} \ldots, \mathbf{x}_{n}, \mathbf{x}_{n+1} \ldots 
\mathbf{x}_{2n}) = \langle \Psi^{\dagger}(\mathbf{x}_{1}) \ldots
\Psi^{\dagger}(\mathbf{x}_{n}) \Psi(\mathbf{x}_{n+1}) \ldots \Psi(\mathbf{x}%
_{2n}) \rangle ,  \label{Eq.Gn}
\end{equation}
where the angle brackets denote the quantum mechanical average. If $G^{(n)}$
factorises 
\begin{equation}
G^{(n)}(\mathbf{x}_{1} \ldots, \mathbf{x}_{2n}) = \langle \Psi^{\dagger}(%
\mathbf{x}_{1}) \rangle \ldots \langle \Psi^{\dagger}(\mathbf{x}_{n})
\rangle \langle \Psi(\mathbf{x}_{n+1}) \rangle \ldots \langle \Psi(\mathbf{x}%
_{n}) \rangle  \label{Eq.Gnfact}
\end{equation}
the field is said to be $n$th order coherent. This particular form of the
coherence functions arose from photodetection theory. The pairing of
creation and annihilation operators is a result of practical photodetectors
measuring photon number. The normal ordering, in which all creation
operators are to the left of all annihilation operators, arises because
detection relies on absorption rather than stimulated emission, which would
be affected by spontaneous emission. An all orders coherent quantum state is
the annihilation operator eigenstate, the so called ``coherent state''. 
This is the quantum state that
most closely approximates (superpositions of) noiseless classical waves.

One difficulty with this definition of coherence is that it relies on the
spontaneous symmetry breaking assumption that the field expectation $\langle
\Psi(\mathbf{x}) \rangle$ value is not zero. Normalised coherence functions
overcome this problem: 
\begin{equation}
g^{(n)}(\mathbf{x}_{1} \ldots, \mathbf{x}_{2n}) = \frac{ G^{(n)}(\mathbf{x}%
_{1} \ldots, \mathbf{x}_{2n}) } {\left( G^{(1)}(\mathbf{x}_{1},\mathbf{x}%
_{1}) \ldots G^{(1)}(\mathbf{x}_{2n}, \mathbf{x}_{2n}) \right)^{1/2} } .
\label{Eq.gn}
\end{equation}
For a maximally coherent field the numerator factorises and $g^{(n)}=1$.

In a simple spatial interference experiment two sources at $x_{1}$ and $%
x_{2} $ propagate for the same time to a detection point. If the sources are
coherent, interference results. The relevant coherence function is
first-order: $g^{(1)}(x_{1},x_{2})$, because the intensity at the detection
point contains the cross term $\langle \Psi^{\dagger}(x_{1})
\Psi(x_{2})\rangle$ in the field amplitudes\cite{WallsMilb}. 
(Note that when either the spatial or
temporal arguments are suppressed, as are the temporal arguments here, they
are assumed to be equal.) If this cross term is zero
the field fluctuations wash out the interference. The visibility of the
interference fringes is maximal when $g^{(1)}(x_{1},x_{2}) = 1$. From the
definition Eq.(\ref{Eq.gn}), this is true for any single mode field, as for $%
g^{(1)}$ the only relevant expectation value is $\langle a^{\dagger} a
\rangle$. In a BEC a large number of atoms are cooled into a single mode.
Therefore a BEC with output coupling that preserves coherence is a prime
candidate for an atom laser.

The first order spatial coherence of two freely expanding Bose-Einstein
condensates has been observed experimentally by the detection of
interference fringes\cite{Andrews97}. In this experiment two condensates
separated by $40 \mu \mbox{m}$ were created in a double-well potential. The
potential was switched off and the condensates left to expand over a period
of $40 \mbox{ms}$ after which time they overlapped. High contrast
matter-wave interference fringes were observed indicating that the BECs were
first order spatially coherent.

Ideally one would want a continuous output beam which is stationary, i.e.
for which the first order temporal coherence, $g^{(1)}(t+\tau,t) =
g^{(1)}(\tau)$ does not depend on $t$. The first order temporal coherence
function for an atom laser beam, at some particular point, is defined as 
\begin{equation}
g^{(1)}(\tau) \equiv \frac{\langle \Psi^{\dag}(t+\tau) \Psi(t) \rangle}{%
\sqrt{\langle \Psi^{\dag}(t) \Psi(t) \rangle} \sqrt{\langle
\Psi^{\dag}(t+\tau) \Psi(t+\tau) \rangle}} . \label{Eq.g1tau} \\
\end{equation}
This is a useful measure of phase
fluctuations, which we require to be small for a laser like source. For a
laserlike, first order temporally coherent source, $g^{(1)}(\tau) \approx 1$
over a time called the coherence time. This is in contrast to a noisy source
for which $g^{(1)}(\tau)$, decays quickly to zero. According to Wiseman\cite%
{Wiseman97} an atom beam might be regarded as laser-like if the coherence
time is much greater than the inverse atomic flux.

The second order coherence function relates two separate field quantum
(atoms or photons) detection events. Unlike the first order coherence
function, it is sensitive to the quantum mechanical state of the field. For
a single mode thermal state $g^{(2)} = 2$ and for a BEC or single mode
coherent state $g^{(2)} = 1$.

The second order spatial coherence of a matter wave source is related to the
atom-atom interaction energy of the condensate\cite{Mewes96}. This is
because for a short-range potential, the interaction energy is proportional
to the probability that two atoms are nearby, which is in turn proportional
to $g^{(2)}$. The experimental evidence\cite{Holland97} is consistent with $%
g^{(2)} = 1.0 \pm 0.2$ for BECs, confirming that condensates suppress local
density fluctuations.

As well as first order temporal coherence, an atom laser beam is expected to
have higher order temporal coherence. The second order temporal coherence
function (for a stationary system) is 
\begin{equation}
g^{(2)}(\tau) \equiv \frac{\langle :I(t+\tau) I(t): \rangle }{\langle I
\rangle ^{2}},  \label{Eq.g2taub}
\end{equation}
where $I=\Psi^{\dagger} \Psi$, and the colons denote normal ordering, i.e.
all creation operators to the left of all annihilation operators. $g^{(2)}$
is approximately unity for a coherent source. A filtered thermal beam, which
is first order coherent, has $g^{(2)}(\tau) = 2$ for short times,
approaching 1 for long times. In terms of counting field quanta $%
g^{(2)}(\tau)$ is determined by the distribution of arrival times. A field
with $g^{(2)}(\tau) = 1$ has arrival times distributed according to
Poissonian statistics, while $g^{(2)}(\tau) > 1$ corresponds to
superpoissonian statistics, or bunching of arrival times. Only purely
quantum mechanical correlations, such as those associated with squeezed
states, allow $g^{(2)}(\tau)$ to be less than one.

Experimental evidence of third order coherence in BECs has been given by
Burt \textit{et al.}\cite{Burt97}. They compared the three-body
recombination rate constant in condensed and non-condensed Bose gases. They
found that the ratio of $g^{(3)}$ for a thermal cloud to that for a BEC was 
$7.4 \pm 2$, in agreement with the predicted value \cite{Kagan85} of $3! = 6$. 
For a BEC $g^{(3)}=1$, and for the non-condensed fraction $g^{(3)}=6$.  It is
notable that $g^{(3)}=1$ is the same as for classical distinguishable
particles.  This is because BEC particles occupy the same field mode
and are therefore already indistinguishable in principle.  Hence state
symmetrization, which is used to enforce indistinguishability, thereby
generating the quantum mechanical factor of $3!$, is neither necessary
nor possible.

For light only the Glauber type coherence functions that we have discussed
are relevant, because all light detectors use photon absorption. However
atoms may be detected in a variety of ways, and different coherence
functions may be measured\cite{Meystre,Naraschewski&Glauber}. For example
the spectrum of off resonant light scattered from a BEC is related to the
density correlation function, rather than the Glauber normally ordered
coherence function\cite{Meystre}. Normally ordered coherence functions are
relevant for destructive detection, such as by ionisation detectors. We have
focussed on them because they allow the most straightforward comparison with
the optical laser.

%%%%%%%%%%%%%%%%%%%%%%%%%%%%%%%%%%%%%%%%%%%%%%%%%%%%%%%%%%%%%%%%%%%%%%%

\section{Atom Laser Models}

\label{sec.AL theory}

In this section we discuss the types of theoretical models which have been
used to describe the atom laser. The starting point is usually an analogy
with the optical laser. However despite the similarities, the differences
between atom and optical lasers are also of considerable interest. We
consider the advantages and disadvantages of three major types of atom laser
models. In order of increasing complexity: rate equation, mean field, and
fully quantum mechanical.

Rate equation models only consider state populations, ignoring the dynamics
of coherences between quantum states. Hence they are unable to properly
describe the quantum mechanical coherence properties of the atom laser.
Nevertheless they usefully describe important physics such as the role of
Bose-enhancement in producing the laser threshold.

The mean-field approach uses Gross-Pitaevskii type equations, Eq.({\ref%
{eq:GP eqn}). Since the atom laser is then described by a macroscopic
wavefunction, quantum coherences may be included in the models. Additional
terms may be introduced to describe the pumping and output coupling. A
limitation of these models is that they assume that the field is in a
coherent state, as discussed at the end of section \ref{sec:Interactions}.
This means that the quantum coherence of the output cannot be calculated.
However the nonlinear dynamics of the BEC and output beam are fully
modelled, as is the spatial propagation of the output beam. }

The most common fully quantum mechanical approach uses the quantum optical
master equation\cite{WallsMilb}. This is an equation for a reduced density
operator, obtained by averaging (strictly tracing) over part of the system,
usually the output and/or pump modes. The Born-Markov approximation is
commonly applied during the trace over the output modes to ensure that a
differential equation, rather than an integro-differential equation, is
obtained\cite{Moy99}. Because this approximation is not necessarily valid we
shall also discuss more general quantum operator models.

In the rest of this section we give brief descriptions of these three
approaches and the kinds of physics they treat.  Pumping and output coupling 
are considered in detail in sections 6 and 7.

\subsection{Rate equations}

\label{subsec:rateeq}

The first rate equation based atom laser models were presented by Olshanii 
\textit{et al.}\cite{Olshanii95} and Spreeuw \textit{et al.}\cite{Spreeuw95}%
. We discuss a later model due to Moy \textit{et al.}\cite{Moy97a} which is
based on implementation in a hollow optical fibre.

\begin{figure}[tbp]
\begin{minipage}[b]{.4\linewidth}
\centering\includegraphics[height=3.5cm]{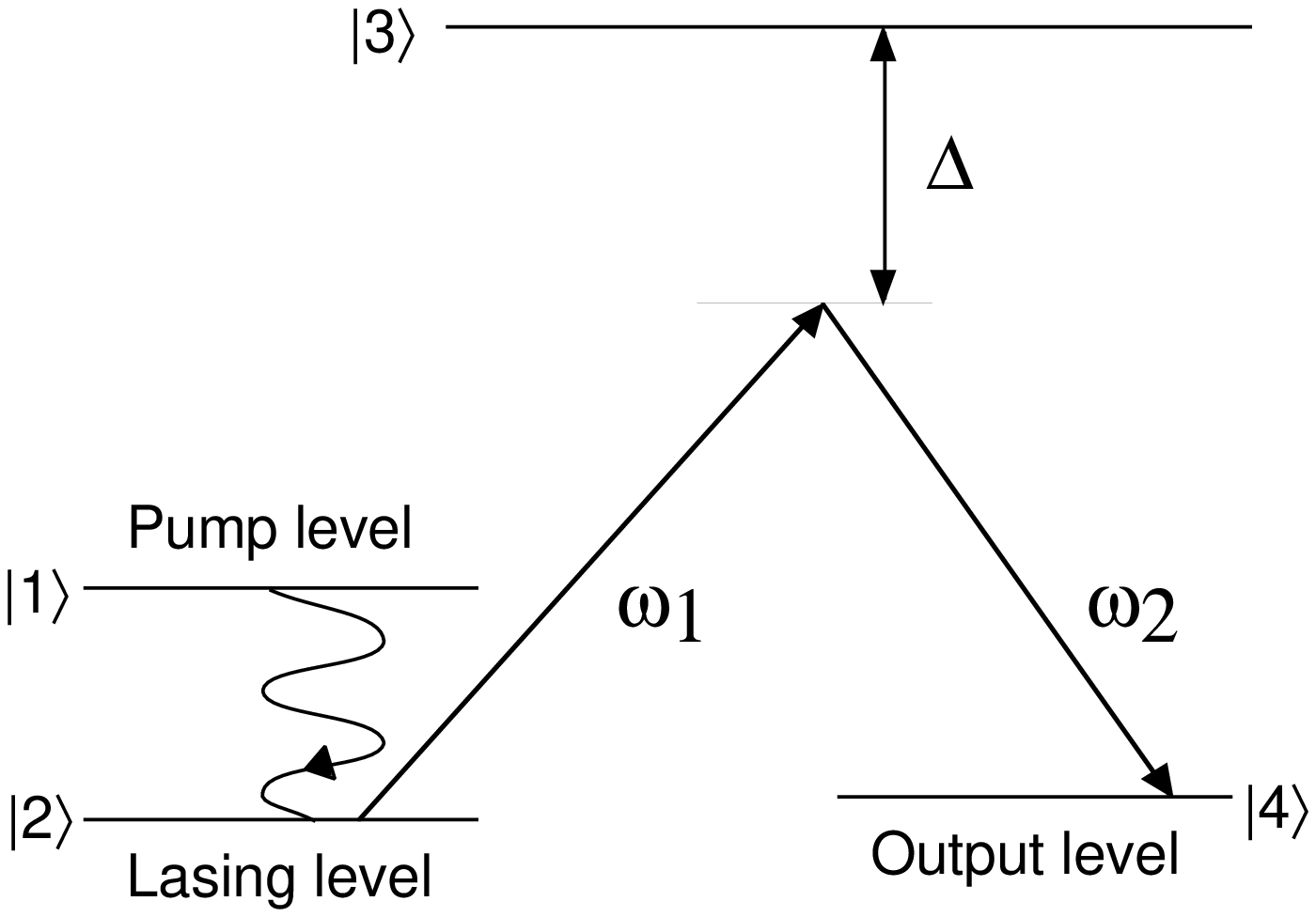}
\vspace{3mm}
\end{minipage}\hfill 
\begin{minipage}[b]{.52\linewidth}
\centering\includegraphics[height=5cm]{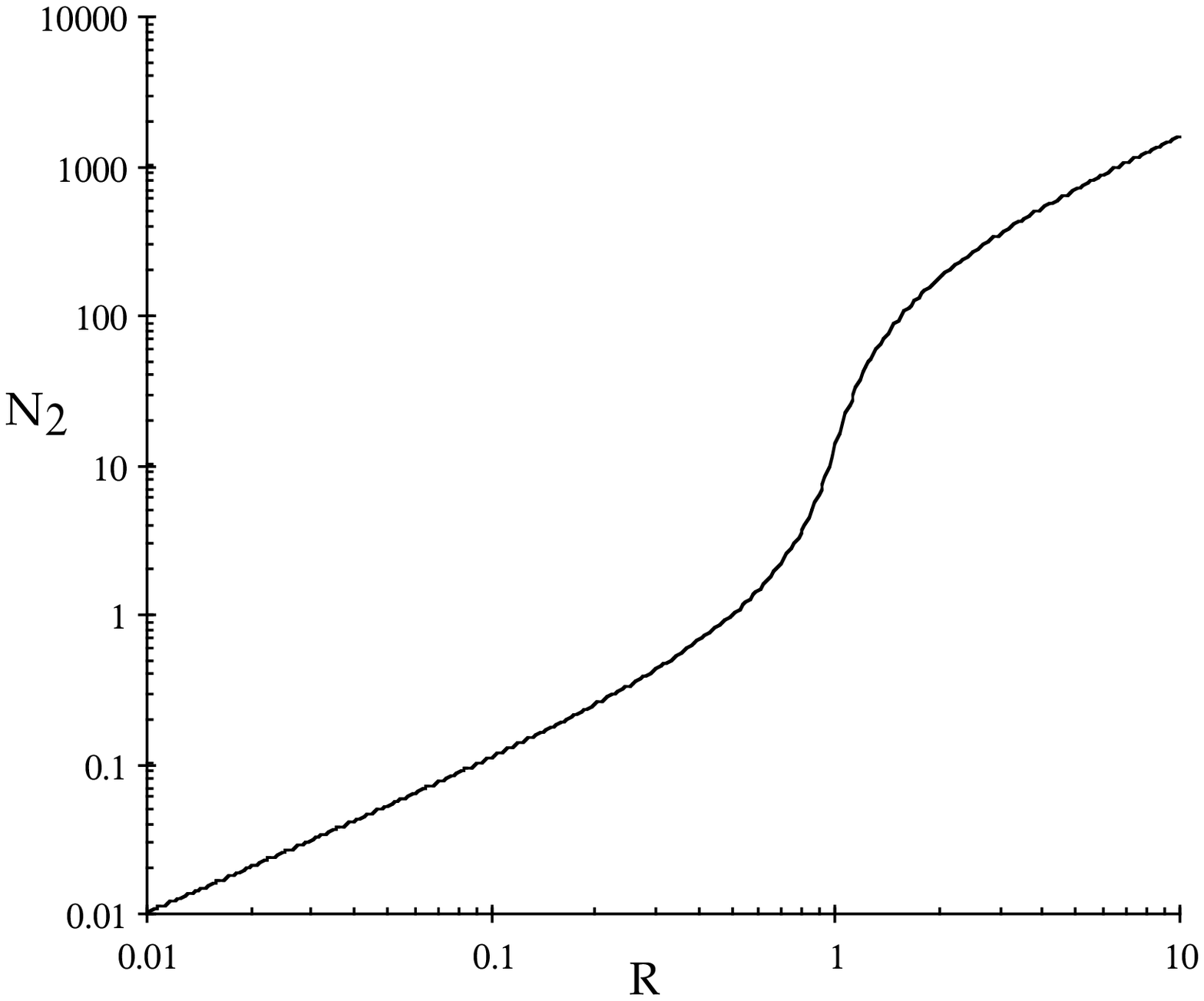}
\end{minipage}\hfill
\caption{Schematic level diagram for the atom laser model of Moy \textit{et
al.}\protect\cite{Moy97a} (left). Plot of the steady state number of atoms
in the lasing mode, as a function of the dimensionless pumping rate, $R$
(right). Threshold occurs at $R=1$.}
\label{fig2:levels and threshold}
\end{figure}

The model consists of atoms with four energy levels, as indicated in Fig.~%
\ref{fig2:levels and threshold}. Level $|1 \rangle$ is the input pump level,
level $|2 \rangle$ is the lasing level and level $|4 \rangle$ is the output
level. Level $|3 \rangle$ mediates the output coupling Raman transition and
is not populated. Level $|1 \rangle$ atoms irreversibly emit a photon to
populate the lasing level $|2 \rangle$ at the Bose-enhanced rate $%
r_{12}(N_{21}+1)$, where $N_{21}$ is the number of atoms in level $|2
\rangle $ and the ground state of the laser trap. A Raman transition couples
atoms from level $|2 \rangle$ into level $|4 \rangle$ with rate constant $%
r_{24}$. 
The resulting rate equations for the numbers of atoms $N_{k}$ in level $|k
\rangle$ are 
\begin{eqnarray}
\frac{d N_{1}}{d t} &=& r_{1} - \sum_{j} g_{j} (N_{2j} + 1) r_{12} N_{1} -
(1-\sum_{j} g_{j}) r_{12} N_{1} ,  \nonumber  \label{dn1dt} \\
\frac{d N_{2j}}{d t} &=& g_{j} r_{12} N_{1} (N_{2j}+1) - r_{24} N_{2j} +
~G_{j} (N_{2j}+1) N_{4} r_{24},  \nonumber \\
\frac{d N_{4}}{d t} &=& \sum_{j} N_{2j} r_{24} -\sum_{j} G_{j} (N_{2j}+1)
N_{4} r_{24} - N_{4} \frac{1}{t_{0}}.  \label{dn4dt}
\end{eqnarray}
The $N_{2j}$ are the numbers of atoms in level $|2 \rangle$ in the $j$th
mode of the lasing level trap. $g_{j}$ is the overlap between the pump mode
and the $j$th mode of the lasing trap. $G_{j} = N_{4j} / N_{4}$ is the
proportion of atoms with electronic level $|4 \rangle$ that have the same
spatial mode as the $j$th mode of the lasing trap. $t_{0}$ is the time scale
on which atoms are irreversibly lost from the system due to the Raman
momentum kick. For a more complete description of these equations refer to
Moy \textit{et al.}\cite{Moy97a}.

The solutions of these rate equations show behaviour familiar from optical
lasers. The steady state population of the laser mode $N_{21}$ as a function
of a suitably scaled pump rate $R=r_{1}g_{1}/r_{24}$ shows a rapid increase
at the threshold value of $R=1$, as shown in Fig.~\ref{fig2:levels and
threshold}. As a function of time the above threshold values of the lasing
cavity mode populations $N_{2j}$ show initial growth, and then all but the
lasing mode $N_{21}$ decay. This is the familiar mode competition in which
the ``winner takes all'' due to runaway Bose stimulation.

The rate equation model of Olshanii \textit{et al.}\cite{Olshanii95} is
similar in form. However they included photon reabsorption and found that
lasing is inhibited by small amounts of reabsorption.

\subsection{Mean-field models}
\label{subsec.MF-models}

In this approach, the atom laser is modelled using the Gross-Pitaevskii (GP)
type equations discussed in section \ref{sec:Interactions}. The
``mean-field'' is the GP wavefunction which determines the effective
potential felt by an atom due to its 2-body interactions with all of the
other atoms in the condensate. Indeed one of the advantages of GP models is
this straightforward treatment of atom-atom interactions.

Kneer \textit{et al.}\cite{Kneer98} describe a GP model of an atom laser
including pump and loss terms. The modified GP equation for the laser mode
macroscopic wavefunction is 
\begin{equation}
i \hbar \frac{\partial \psi}{\partial t} = - \frac{\hbar^{2}}{2 m} \frac{%
\partial^{2} \psi}{\partial x^{2}}  + V_{E}(r) \psi + U_{0}|\psi|^{2}
\psi ~ + H_{\mbox{\small gain}}\psi + H_{\mbox{\small
loss}}\psi .  \label{Eq.PumpDampGP}
\end{equation}
The first three terms are standard. The gain and loss are represented by
phenomenological terms with rate constants $\Gamma$ and $\gamma_{c}$
respectively, 
\begin{equation}
H_{\mbox{\small gain}}\psi = \frac{i \hbar}{2} \Gamma N_{u} \psi, \quad H_{%
\mbox{\small loss}}\psi = -\frac{i \hbar}{2} \gamma_{c} \psi.
\label{Eq.Hloss}
\end{equation}
In addition to the GP equation (\ref{Eq.PumpDampGP}), a rate equation is
used to describe the number of uncondensed atoms, $N_{u}$, 
\begin{eqnarray}
\frac{d N_{u}}{dt} &=& R_{u} - \gamma_{u} N_{u} - \Gamma N_{c} N_{u}.
\label{Eq.Nu}
\end{eqnarray}
$R_{u}$ describes a source which pumps the uncondensed state at rate $R_{u}$%
. The term $\gamma_{u} N_{u}$ describes atoms lost from the system, but not
trapped in the condensed state. The final term describes the transfer of
atoms into the condensed state. The condensate loss is spatially
homogeneous, and undamped collective excitations are found. To avoid this
they introduce a spatially dependent decay rate, so that $\gamma_{c}$ is a
function of position, $\gamma_{c}(x)$. This produces a single lasing mode as
the spatially dependent loss leads to mode selection.

Robins \textit{et al.}\cite{Robins00} present a GP model which combines the
pumping model of Kneer \textit{et al.}, with an explicit treatment of the
output coupling and of the output beam. Including the output beam in the
model allows one to investigate its spectrum and spatial distribution.
Furthermore, non-Markovian effects in the output, such as Rabi oscillations
between the condensate and beam, can be explored. The equations which
describe the laser mode $\psi^a (x,t)$, and the output beam $\psi^b (x,t)$ are 
\begin{eqnarray}
i\hbar \frac{\partial\psi^a}{\partial t} &=& -\frac{\hbar^2}{2m}\frac{%
\partial^2}{\partial x^2}\psi^a + \frac{1}{2}m\omega^2x^2\psi^a +
U^a_0|\psi^a|^2\psi^a + U^a_0|\psi^b|^2\psi^a  \nonumber \\
&& -i \gamma_{r} ( |\psi^a|^4 + |\psi^b|^4) \psi^a - \hbar\Gamma_{R}
e^{ik_0x}\psi^b(x) + \frac{i \hbar}{2}~ \Gamma ~N_{u} \psi^{a} ,  \nonumber
\\
i\hbar \frac{\partial\psi^b}{\partial t} &=& -\frac{\hbar^2}{2m}\frac{%
\partial^2}{\partial x^2}\psi^b +mgx \psi^b+ U^b_0|\psi^b|^2\psi^b +
U^a_0|\psi^a|^2\psi^b  \nonumber \\
&& -i \gamma_{r} ( |\psi^a|^4 + |\psi^b|^4) \psi^b - \hbar\Gamma_{R}
e^{-ik_0x}\psi^a .  \label{Eq.Nurate2}
\end{eqnarray}
\begin{figure}[tbp]
\includegraphics[width=\textwidth]{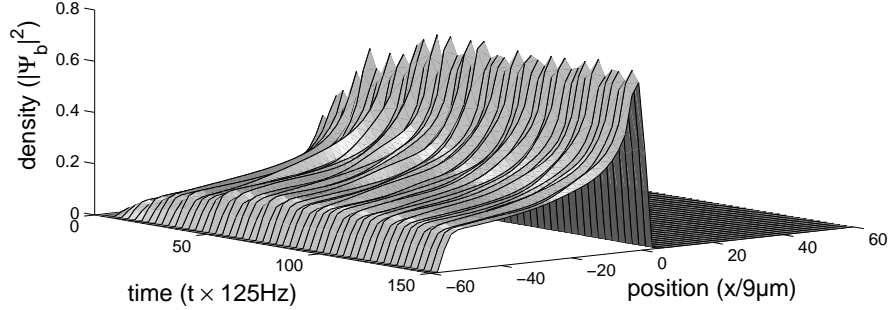}
\caption{Atom laser beam density as a function of position and time, showing
a quasi-steady state. Atoms are absorbed at the spatial boundaries.
Parameters are given by Robins \textit{et al.}\protect\cite{Robins00}.}
\label{fig3:GP output}
\end{figure}
The number of uncondensed atoms $N_{u}$ obeys Eq.(\ref{Eq.Nu}).  The
Raman output coupling is described by the terms containing the
momentum kick generators $e^{\pm ik_0x}$.  Three body recombination
has been included as a condensate loss mechanism, by the terms
proportional to $\gamma_{r}$.  This is found to have a stabilising
effect on the non-linear dynamics of the atom laser.  Two-body losses
have not been included.

Typical output from a one dimensional numerical solution of this model
is shown in Fig.~\ref{fig3:GP output}. 
The oscillations represent nonlinear dynamical noise,
which will determine the laser's effective linewidth, or first order
coherence. This is an example of physics which cannot be treated by the rate
equation approach.

How mean field models may be extended to incorporate finite
temperature effects, such as the population of excited levels, is
described in subsequent sections, and by Burnett and Griffin in this
volume.

\subsection{Fully quantum mechanical models}

A number of quantum optical master equation based models of the atom laser
have been discussed in the literature\cite%
{Holland96,Guzman96,Wiseman95a,Wiseman96}. The model of Holland \textit{et
al.}\cite{Holland96} is based on evaporative cooling and incorporates: the
creation of an atom in the trap from the pump field, the loss of an atom
from a high lying trap state, collisions which scatter atoms into or out of
the ground state, and the loss of atoms from the ground state through output
coupling.

A three level model is sufficient to describe these features. Atoms are
injected with rate $\kappa_{1}$ in level one and are lost with rates $%
\kappa_{0}$ and $\kappa_{2}$ from level zero and level two respectively.
Only the energy conserving 2-body interaction terms are considered. Level
two is adiabatically eliminated by assuming large damping due to the
evaporative cooling. This leads to the master equation for the density
operator $\rho$, 
\begin{eqnarray}
\frac{\partial \rho}{\partial t} &=& \frac{1}{i \hbar} \left[V_{i}, \rho %
\right] ~ + \kappa_{0} \mathcal{D}[a_{i}]\rho + \kappa_{1} \mathcal{D}%
[a_{i}^{\dag}]\rho + \Omega_{r}\mathcal{D}[a_{0}^{\dag} a_{1}^{2}] \rho
\end{eqnarray}
where 
\begin{equation}
V_{i} = \hbar (a_{0}^{\dag} a_{0} - 2 n_{0}) (V_{0000} a_{0}^{\dag} a_{0} +
V_{0101} a_{1}^{\dag} a_{1}) + \hbar V_{1111} a_{1}^{\dag 2} a_{1}^{2},
\end{equation}
and $n_{0}$ is the mean population of level zero. The loss terms are assumed
to be of the standard (Born-Markovian) Lindblad form, defined by 
\begin{equation}
\mathcal{D}[c] \rho = 2 c \rho c^{\dag} - (c^{\dag} c \rho + \rho c^{\dag} c)
\label{Eq.A2}
\end{equation}
The effective redistribution rate $\Omega_{r}$ is proportional to $%
|V_{0211}|^{2} / \kappa_{2}$.

Equations for the populations, $\langle n k | \rho(t) | n k \rangle$, where $%
|n k \rangle$ describes a state with $n$ atoms in level zero and $k=0$ or $%
k=1$ atoms in level one, are obtained. Level one is rapidly depleted by two
atoms scattering into levels zero and two. The steady state population of
the level zero lasing level is found to be Poissonian with mean atom number $%
2 \kappa_{1}/(3 \kappa_{0})$. Thus when loss out of the lasing mode is
sufficiently small $\kappa_{0} << \kappa_{1}$, a large population can build
up in the lasing mode.

In order to produce a state with a well defined phase they consider
parameter regimes where $\Omega_{r}$ is much larger than the frequency
shifts in $V_{i}$. This inhibits the dispersive nature of the latter from
destroying phase coherence. They find that, as for an optical laser, line
narrowing occurs as the population of the ground state increases. However,
in a refinement of the model due to Wiseman \textit{et al.}\cite{Wiseman96}
the linewidth of the laser output is broader than the bare linewidth of the
laser mode.

The crucial refinement is in the description of the pumping of level 1. The
approximation that Holland \textit{et al.} make allows atoms to leak into,
but not out of, the trap. To allow the latter Wiseman \textit{et al.}
replace the term $\kappa_{1} \mathcal{D}[a_{1}^{\dag}] \rho$ with 
\begin{equation}
\kappa_{1} (N+1) \mathcal{D}[a_{1}^{\dag}]\rho + \kappa_{1} N \mathcal{D}%
[a_{1}]\rho.
\end{equation}
The $N$ is the mean population of the pump reservoir modes. In fact, there
is no limit in which the extra terms can be ignored. When $\kappa_{0} <<
\kappa_{1}$, there are two distinct parameter regimes depending on the
relative size of a parameter proportional to $\kappa_{2} |V_{0211}|^{2}$,
which they call $\Gamma$, and $\kappa_{0}$. These they call the strong and
weak collision regimes: $\Gamma >> \kappa_{0}$ and $\Gamma << \kappa_{0}$
respectively. In the weak collision regime, $\Gamma << \kappa_{0}$ there is
a large population in the laser mode if $N >> 1$, i.e. with strong pumping.
The power spectrum reveals a linewidth larger than the bare linewidth of the
lasing mode. Nevertheless, the phase diffusion rate may still be slow in the
sense of being much less than the output flux of atoms from the lasing mode.

Wiseman and Collett\cite{Wiseman95} proposed an atom laser model using
optical dark state cooling. The laser is found to have a linewidth narrower
than the bare linewidth. This is an improvement on models based on
evaporative cooling discussed above. The output is also found to be second
order coherent.

The master equation is one means of describing the full quantum statistics
of an atom laser. It has the great advantage of drawing on the wealth of
techniques that have been developed in quantum optics\cite{WallsMilb}.
However, a fully quantum theory of the atom laser can be developed without
using the standard master equation approximations\cite{Hope97,Moy97b,Hope00}%
. In particular, assuming a Born-Markov form for the output coupling ignores
the possibility of the output beam acting back on the laser mode. This may
happen, for example, due to Rabi cycling between the output and laser modes.

Hope\cite{Hope97} developed a general theory of the input and output of
atoms from an atomic trap. Moy \textit{et al.}\cite{Moy97b} used this theory
to find an analytic expression for the spectrum of a beam of atoms output
from a single mode atomic cavity. For very strong coupling it is just the
spectrum of the trapped lasing mode, and is therefore broad. As the strength
of the coupling is reduced, the linewidth is reduced. For small coupling
strengths the lineshape approaches a Lorentzian, centred about the energy of
the cavity with a linewidth proportional to the coupling strength. Hope 
\textit{et al.}\cite{Hope00} include a pumping mechanism and they find that
the output spectrum is substantially wider than that obtained using the
Born-Markov approximation. However an important limitation of these theories
is that they do not include 2-body interactions.

%%%%%%RJB%%%%%%%%%%%%%%%%%%%%%%%%%%%%%%%%%%%%%%%%%%%%%%%%%%%%%%%%%%%%%%%%%%%%%

\section{Atomic Gain}

\label{Sect.Gain}

We have seen in sections \ref{sec.Introduction} and \ref{sec.AL theory} that
a mechanism for atomic gain, to provide coherent amplification of the atomic
field in the lasing mode, is an essential element of an atom laser. A number
of gain mechanisms have been proposed, (e.g. see section \ref{subsec:rateeq})
but to date only one, evaporative cooling, has been demonstrated
experimentally. The process of evaporative cooling is based on the removal
of high energy atoms from the gas, so that subsequent collisional
rethermalisation of the remaining atoms results in a distribution of lower
temperature (i.e. lower average energy). Under appropriate conditions,
bosonic stimulation of transitions into the ground state of the confining
potential leads to the formation of a Bose condensate: in an atom laser this
would constitute the lasing mode.

The kinetics of condensate growth has long been a subject of theoretical
study$^{47-49}$,
%\cite{GrowthGeneral,GrowthStoof,Kagan} 
but the first quantitative
calculations were carried out only recently by Gardiner. Together with coworkers he has
given a comprehensive treatment of condensate formation by bosonic
stimulation, in a series of papers$^{50-55}$.
%\cite{QK1,bosgro,QK3,newbosgro,QK5,QK6}.
In this section we review the major features and results of this theory.

\subsection{Quantum Kinetic theory}

\begin{figure}[tbp]
\centering \includegraphics[width=6cm]{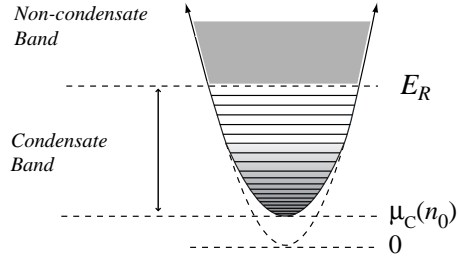}
\caption{Schematic representation of modification of trap energy levels by
condensate meanfield. Levels above $E_{R}$ are unaffected.}
\label{Fig.levelschange}
\end{figure}

The starting point of the treatment is the third in a series of papers 
on Quantum Kinetic theory developed
by Gardiner and Zoller\cite{QK3}, which we shall refer to as QKIII. We note
that an alternative formulation of a Quantum Kinetic theory has been given by
Griffin and coworkers (see this volume, and references therein). The gas of
Bose atoms in a harmonic trap of frequency $\omega $ is described by a
second-quantized field, as in Eqs.(\ref{eq:hamiltonian} - \ref{eq:2nd
quantised H}).  Two-body collisions are included but three-body collisions are
neglected. Guided by the fact that the energy spectrum above a certain value 
$E_{R}$ is unaffected by the presence of the condensate, we can divide the
field into the condensate band (all energy levels below $E_{R}$) and the
non-condensate band (all the energy levels above $E_{R}$), as represented
in Fig. \ref{Fig.levelschange}.

The noncondensate band (which contains the majority of the atoms) is taken
to be fully thermalised, in the sense that for any point $\mathbf{r,k}$ in
phase space the single particle distribution function $F(\mathbf{r,k})$ (in
principle a Wigner function) is localised, and has a well defined
temperature $T$ and chemical potential $\mu $. Thus the noncondensate band
can be treated as a time independent bath, and a master equation is derived
for the density matrix of the condensate band, (Eqs. (50a)-(50f) QKIII). It
is worth emphasizing that in general the condensate band is not in thermal
equilibrium, and the master equation provides a description of its dynamical
evolution. The level structure of the condensate band is well described by
the Bogoliubov spectrum, and Gardiner\cite{truebog} has shown that
appropriate states for this band are $|N,\{n_{m}\}\rangle $ which have a
definite number $N$ of total particles in the band, and a set of numbers $%
\{n_{m}\}$ representing the occupation numbers of the condensate ($m=0$)
and quasiparticle (excited Bogoliubov) levels. In this treatment, the
condensate wavefunction and energy eigenvalue (the condensate chemical
potential $\mu _{C}(N)$ ) are given by the time-independent Gross-Pitaevskii
equation with $n_{0}$ atoms. The quasiparticles may be particle-like
excitations (at higher energies) or phonon-like collective excitations (at
low energies). The phonon excitations are not eigenstates of the number
operator and thus the total population of the quasiparticles need not be
conserved, in contrast to $N\;$which is strictly conserved.

By taking a representation of the master equation in terms of the states $%
|N,\{n_{m}\}\rangle $, and then taking the diagonal elements, a stochastic
master equation for the occupation probabilities $p(N,\{n_{m}\})$ is
obtained. The transitions appearing in this equation can be divided into
those which change the number of atoms $N$ in the condensate band, and those
which do not. The former, which we call \textit{growth processes}, arise
predominantly from collisions of two noncondensate atoms where one of the
outgoing atoms goes into the condensate band. The reverse collision is also
included in the growth processes. The transitions which do not change $N$
are called \textit{scattering processes,} and arise predominantly from
collisions between a condensate band atom and a noncondensate band atom
which leave one atom in each band. Other collision processes, in which the
pair of colliding atoms either begins or ends in the condensate band, will
in principle also contribute to growth or scattering, but can be neglected
since the condensate band contains such a small fraction of all the atoms.
We note that collisions between pairs of atoms that begin and remain in the
noncondensate band play the role of thermalising the bath and do not appear
in this master equation.

\subsection{\protect `Particle-like' rate equations}

\label{Sect.rate_eqns}

Using standard methods, a deterministic rate equation for the mean
population $\langle n_{m}\rangle $ of the $m$th quasiparticle level (which
we will subsequently write $n_{m}$) can be obtained from the master equation
and we write it as 
\begin{equation}
\frac{dn_{m}}{dt}=\dot{n}_{m}|_{\mathrm{growth}}+\dot{n}_{m}|_{\mathrm{scatt}%
}  \label{Eq.qp_rate}
\end{equation}%
where the forms for the growth transition rates $\dot{n}_{m}|_{\mathrm{growth%
}}$ and the scattering transition rates $\dot{n}_{m}|_{\mathrm{scatt}}$ will
be discussed below. A\ similar deterministic rate equation can be written
for the mean population of the condensate band $\langle N\rangle $ (which we
will subsequently write $N$).

The growth transition rates $\dot{n}_{m}|_{\mathrm{growth}}$ can be obtained
in explicit form from the population master equation given in Eq.(189)
QKIII, and comprise all the processes where $N\;$changes by $1,$ and one of
the quasiparticle populations, $n_{j}$ say, changes by $1$ or $0,$ (with all
other $n_{k}$ unchanged). Most of these processes (e.g. $N$ $\rightarrow N+1$
and $n_{j}\rightarrow n_{j}+1,$ or $N$ $\rightarrow N+1$ and $\{n_{m}\}$
unchanged) admit a particle-like interpretation. However, those processes
where $N$ and $n_{j}$ change in opposite directions have to be interpreted
as phonon-like processes, and for example could represent a process where
the condensate is initially oscillating, and a thermal atom is added to the
condensate band and causes the condensate oscillation to be reduced. All
such phonon-like processes are neglected in this treatment, since only a
small fraction of quasiparticle wavefunctions have phonon-like character,
and we expect they will play an unimportant part in the condensate growth.
The growth terms for the quasiparticle levels thus simplify to the
form\smallskip 
\begin{equation}
\dot{n}_{m}|_{\mathrm{growth}}=2W_{m}^{++}(N)\left( n_{m}+1\right)
-2W_{m}^{--}(N)n_{m}  \label{Eq.growth}
\end{equation}%
and the equation for the mean number in the condensate band becomes 
\begin{equation}
\frac{dN}{dt}=2W^{+}(N)\left( N+1\right) -2W^{-}(N)N
\label{eq:simple_growth}
\end{equation}%
where $W_{m}^{++}(N)$ and $W^{+}(N)$ are forward rates and $W_{m}^{--}(N)$
and $W^{-}(N)$ are backward rates. It is shown in QKIII that when the
condensate wavefunction $\xi _{N}(\mathbf{r})$ is sharply peaked compared to
the spatial dependence of $F(\mathbf{r,k}),$ then 
\begin{eqnarray}
W^{+}(N) &=&{\frac{4a^{2}}{m^{2}(2\pi )^{3}}}\int d^{3}\mathbf{K}_{1}\int
d^{3}\mathbf{K}_{2}\int d^{3}\mathbf{K}_{3}\int d^{3}\mathbf{k}\,\delta (%
\mathbf{K}_{1}+\mathbf{K}_{2}-\mathbf{K}_{3}-\mathbf{k})  \nonumber \\
&&\!\!\!\!\!\!\!\delta (\Delta \omega _{123}-\mu _{C}(N)/\hbar )F(0\mathbf{,K%
}_{1})F(0\mathbf{,K}_{2})\left( 1+F(0\mathbf{,K}_{3})\right) |\tilde{\xi}%
_{N}(\mathbf{k})|^{2}  \label{Eq.Wplus}
\end{eqnarray}%
\newline
The integrand represents a collision in which a pair of noncondensate atoms
of \noindent momenta $\mathbf{K}_{1}$ and $\mathbf{K}_{2}$ produces one
noncondensate atom ($\mathbf{K}_{3}$) and one condensate band atom of
momentum $\mathbf{k}$. Energy conservation is expressed in the second $%
\delta $ function, in which $\Delta \omega _{123}=\omega _{\mathbf{K}%
_{1}}+\omega _{\mathbf{K}_{2}}-\omega _{\mathbf{K}_{3}},$ where $\hbar
\omega _{\mathbf{K}_{i}}=$ $\hbar ^{2}\mathbf{K}_{i}^{2}/2m$ $+V_{T}(0)$ is
the particle energy at the centre of the trap potential $V_{T}$. $\tilde{\xi}%
_{N}(\mathbf{k})$ is the momentum-space ground-state wavefunction. The
expression for $W_{m}^{++}(N)$ is similar to Eq.(\ref{Eq.Wplus}), and the
forward and backward rates are related by 
\begin{eqnarray}
W^{+}(N) &=&e^{[\mu -\mu _{C}(N)]/kT}W^{-}(N)  \label{Eq.Wplusreverse} \\
W_{m}^{++}(N) &=&e^{[\mu -e_{m}]/kT}W_{m}^{--}(N)  \label{Eq.Wmppreverse}
\end{eqnarray}%
where $e_{m}$ is the energy of the $m^{th}$ quasiparticle level. Equations (%
\ref{Eq.qp_rate})-(\ref{Eq.Wmppreverse}) are derived within the
approximation that the number of particles in the condensate, $n_{0},$ is
large enough that we can write $n_{0}\approx N,($which is required for the
Bogoliubov spectrum to be accurate). However, in the early stages of
condensate growth when $n_{0}$ is small, the unperturbed spectrum should be
used, and will be accurate provided none of the condensate band levels are
too highly populated. This means that by making the replacement $%
N\rightarrow n_{0}$ in Eqs.(\ref{Eq.qp_rate})-(\ref{Eq.Wmppreverse}) those
equations will provide a good description of the dynamical evolution of the
levels in the condensate band throughout the formation of the condensate.
Values for the rates $W^{+}$ can be obtained in varying levels of
approximation, and the simplest expression, used in the first quantitative
description of growth\cite{bosgro}, is obtained by assuming a classical
Maxwell-Boltzmann distribution for $F(\mathbf{r,k}),$ and allowing the
integrals in $\mathbf{K}_{1,2,3}$ in Eq.(\ref{Eq.Wplus}) to extend over all
energies, rather than just the noncondensate band. This leads to the
explicit expression 
\begin{equation}
W^{+}(n_{0})\approx {\frac{4m(akT)^{2}}{\pi \hbar ^{3}}}e^{2\mu /kT}
\label{Eq.WplusMB}
\end{equation}%
in which we note the prefactor to the exponential is independent of $n_{0}$
and is essentially the elastic collision rate $\rho \sigma v$ with
quantities evaluated at the critical point for condensation. More accurate
evaluations\cite{newbosgro,QK6} of $W^{+}$ which take into account the
Bose-Einstein nature of the non-condensate distribution and exclude the
condensate region in the $\mathbf{K}$ integrals, give a value about three
times larger than that of Eq.(\ref{Eq.WplusMB}). $W_{m}^{++}(n_{0}),$ which
has a form similar to Eq.(\ref{Eq.Wplus}), is more difficult to evaluate but
is of the same order of magnitude as $W^{+}(n_{0}),$ and in the first
instance we set $W_{m}^{++}(n_{0})\approx W^{+}(n_{0}).$ The sensitivity of
the results to this approximation will be discussed in section \ref%
{Sect.Growth_results}. Finally, in order to reduce the number of equations
to a tractable set, an ergodic assumption is made which allows us to group
together levels of similar energy into small sub-bands, with only the ground
state being described by a single level. The sub-bands have (central) energy 
$e_{m}$ and degeneracy $g_{m}$, and the equations for the growth processes
of these bands are 
\begin{eqnarray}
\dot{n}_{m}|_{\mathrm{growth}} &=&2W_{m}^{++}(n_{0})\left\{ \left[
1-e^{[e_{m}-\mu ]/kT}\right] n_{m}+g_{m}\right\} ,
\label{Eq.growth_explicit} \\
\dot{n}_{0}|_{\mathrm{growth}} &=&2W^{+}(n_{0})\left\{ \left[ 1-e^{[\mu
_{C}(n_{0})-\mu ]/kT}\right] n_{0}+1\right\} .  \label{Eq.growth_n0}
\end{eqnarray}%
These equations exhibit the most important features of condensate growth, or
atomic gain, as we shall discuss shortly in section \ref{Sect.Growth_results}%
. The remaining terms required for Eq.(\ref{Eq.qp_rate}), the scattering
terms $\dot{n}_{m}|_{\mathrm{scatt}}$, describe processes in which a
condensate band atom and a noncondensate band atom collide and leave one
atom in each band. Gardiner and Lee\cite{QK6} have evaluated the scattering
terms within well defined approximations (for the case of an isotropic
trap), to give 
\begin{eqnarray}
&&\dot{n}_{m}|_{\mathrm{scatt}}={\frac{8ma^{2}\omega ^{2}}{\pi \hbar }}%
e^{\mu /k_{B}T}\Gamma (T)\times  \nonumber \\
&&\,\,\Bigg\{\sum_{k<m}{\frac{1}{g_{m}}}\left[ n_{k}(g_{m}+n_{m}){e^{-\hbar
\omega _{mk}/k_{B}T}}-n_{m}(g_{k}+n_{k})\right]  \nonumber \\
&&\,\,+\sum_{k>m}{\frac{1}{g_{k}}}\left[
n_{k}(g_{m}+n_{m})-n_{m}(g_{k}+n_{k}){e^{-\hbar \omega _{km}/k_{B}T}}\right] %
\Bigg\}.  \label{Eq.explicit_scatt}
\end{eqnarray}%
where $\Gamma (T)=\sum_{e_{m}>E_{R}}e^{-e_{m}/kT}$ , $\hbar \omega
_{nq}=e_{n}-e_{q}$, and the notation $k>m$ means $e_{k}>e_{m}$.

\subsection{Results}

\label{Sect.Growth_results}

Equations (\ref{Eq.growth_explicit} - \ref{Eq.explicit_scatt}) provide an
explicit form for the set of deterministic rate equations for populations in
the condensate band, with the size of the set determined by the amount of
level grouping. Much of the behaviour that occurs in the full set of
equations is captured by a simple implementation in which the condensate
band is represented by only one level\cite{bosgro}, i.e. the condensate. The
equations then reduce to a single equation (\ref{Eq.growth_n0}), which has
the laser-like features of stimulated transition rates (the terms
proportional to $n_{0}$) and a spontaneous transition rate (the term +1 in
the curly braces) which initiates growth. In addition, this so-called 
\textit{simple growth equation} has thermodynamic features, in that the
forward and backward transition rates become equal (i.e. equilibrium is
reached) when the chemical potential of the condensate, $\mu _{C}(n_{0}),$
becomes equal to $\mu $ of the thermal bath (to order $1/N$). A crucial
aspect of the growth process is that as the condensate grows, the meanfield
causes $\mu _{C}(n_{0})$ to increase, beginning from $\mu _{C}(0)=$ $3\hbar
\omega /2$ (the trap zero-point energy), and rising as $\left( n_{0}\right)
^{2/5}$ (in the Thomas-Fermi approximation) at large $n_{0}$. It is clear
that a condensate can only grow if the thermal bath has $\mu >0,$ and this
is achieved in the process of evaporative cooling by removing the highest
energy atoms. The higher energy levels quickly come to their equilibrium
distributions, but the lower energy levels remain far from equilibrium, and
the ensuing tendency of collisions of lower energy bath atoms to feed the
condensate band can be characterised by a bath chemical potential $\mu >0.$

\begin{figure}[tbp]
\centering  \includegraphics[width=7.5cm]{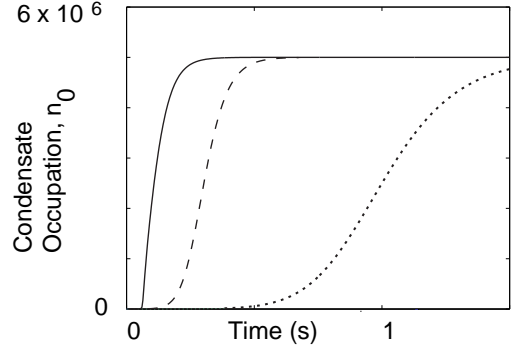}
\caption{Typical condensate growth curves for different implementations of
the model: simple growth equation (dotted line); full model (solid line);
with no scattering (dashed line). Parameters are appropriate to the regime
of the MIT\ experiment: $\protect\mu =43.3\hbar \protect\omega $, $T=900$
nK, $n_{0}(t=0)=100.$ }
\label{Fig.Gain_simple}
\end{figure}

A typical result for the simple growth equation is shown as the dotted line
in Fig.\ref{Fig.Gain_simple} 
where we have chosen a case corresponding to the conditions of the MIT
growth experiment\cite{MITgrowth} in which sodium atoms of scattering length 
$a=2.75$ nm are contained in a trap of geometrical mean frequency $\omega
=2\pi \times 50$ Hz. Although we do not expect quantitative accuracy from
this simple model, the characteristic S shape of the growth curve exhibits
the key aspects of the growth process. In the initial stages the growth is
slow and arises from spontaneous transitions; then as $n_{0}$ increases,
stimulated transitions dominate and the growth is steep. Finally, as the
condensate chemical potential approaches $\mu ,$ there is a slow approach to
equilibrium.

Of course for a quantitatively accurate treatment, more condensate band
levels must be included, along with the more careful calculation for $W^{+}$
(see Eq.(10) ref\cite{newbosgro}). Typically 20-50 sub-bands are used, each
of width $\approx $ $\hbar \omega .$ The energy $e_{m}$ of these sub-bands
increase as $n_{0}$ grows, with the lower levels most affected, and the
levels at the top of the condensate band not affected at all. Although the
increase of the condensate energy level is crucial to the growth process and
must be treated as accurately as possible, it is found that there is little
sensitivity to the scheme used to alter the quasiparticle energy levels.
Typically, the levels between $\mu _{C}(n_{0})$ and $2\mu _{C}(n_{0})$ are
compressed from the bottom in a linear fashion, and higher levels are
unaltered. The solid line in Fig.\ref{Fig.Gain_simple} shows the effect of
using a set of sub-bands in the growth process. The physical parameters are
identical to those used in the simple growth model (the dotted curve) but
there has been a speed up of about an order of magnitude. The effects of the
scattering terms can be separated from the growth terms, as we show with the
dashed line in Fig.\ref{Fig.Gain_simple}, which is a calculation using the
same number of sub-bands as for the solid curve, but with all scattering
terms put to zero. It is clear that the main effect of the scattering
processes is to cause the initiation of condensate growth to occur much more
sharply. This is due to the fact that the growth processes can initially
populate a wide number of sub-bands, and then scattering allows these
populations to be redistributed directly into the condensate. The steep part
of the growth curve is essentially unchanged by scattering, since it is
dominated by the stimulated terms in the growth processes.

A number of sensitivity studies were carried out by Gardiner and Lee\cite%
{QK6}, who showed that the scattering transition rates could be varied by
two orders of magnitude, the relative size of $W^{++}$ to $W^{+}$ could be
varied by a factor of 2, and a wide variety of initial conditions for the
rate equations could be used, with only moderate effects on the growth
curves.

\subsection{\protect Full dynamical solution}

The master equation approach described above, which assumes a nondepletable
bath of fixed thermodynamic properties, can be expected to be inaccurate
when high fraction condensates form. Under those circumstances the full
dynamics of the system must be considered, and Davis {\it et al.} \cite{QK7} have
treated this problem using a modified Quantum Boltzmann formulation. In
their treatment the single particle distribution function $f_{n}$ for the $%
n^{th}$ energy sub-band evolves according to 
\begin{eqnarray}
g_{n}\frac{\partial f_{n}}{\partial t} &=&\frac{8ma^{2}\omega ^{2}}{\pi
\hbar }\sum_{p,q,r}\delta _{e_{r}+e_{n},e_{p}+e_{q}}g_{\min (p,q,r,n)} 
\nonumber \\
&&\times \left\{ f_{p}f_{q}\left( 1+f_{r}\right) \left( 1+f_{n}\right)
-\left( 1+f_{p}\right) \left( 1+f_{q}\right) f_{r}f_{n}\right\} .
\label{Eq:MQBE}
\end{eqnarray}%
This equation appears similar to the ergodic form of the
well-known Quantum Boltzmann equation (e.g.~see ref \cite{HollandQMB}) but
the crucial modification from Quantum Kinetic theory is that the energy
levels $e_{m}$ and the degeneracies $g_{m}$ alter as the condensate
population changes. A\ division of the levels into condensate and
noncondensate bands is now made only for computational purposes, and the
evolution of all sub-band levels is explicitly followed. In addition to the
collisional processes treated in the master equation method, Eq.(\ref%
{Eq:MQBE}) includes thermal scattering processes where the pair of colliding
atoms begin and remain in the noncondensate band, as well as collisions
which either begin or end with a pair of atoms in the condensate band. As
before, all levels are treated as particle-like, which allows an analytic
expression to be obtained for the density of states and its dependence on
the condensate occupation. The numerical simulation of Eq.(\ref{Eq:MQBE}%
), which we shall refer to as the \textit{full dynamical solution}, requires
vastly more computation than the rate equation model described in section %
\ref{Sect.rate_eqns}, and considerable care must be taken with ensuring
energy conservation in the collisional calculations and in rebinning the
sub-bands at each time step. Details are given in ref~\cite{QK7}.

\begin{figure}[tbp]
\centering  \includegraphics[width=\textwidth]{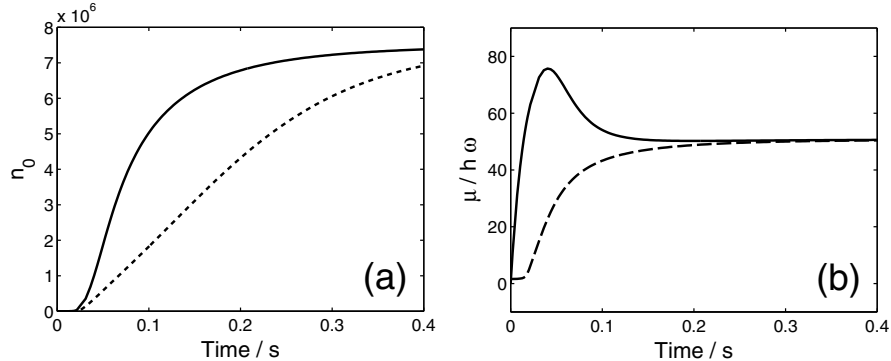}
\caption{(a) Comparison of condensate growth models for a final condensate
fraction of 24\%. Solid line, simulation of Eq.(\ref{Eq:MQBE}) [with $%
\protect\mu =0$ before evaporative cut]. Dotted line, rate equation model.
Final condensate parameters $n_{0}=7.5\times 10^{6},T=590$nK. (b) Chemical
potentials $\protect\mu _{\mathrm{eff}}$ (solid line) and $\protect\mu %
_{C}(n_{0})$ (dashed line).}
\label{Fig.MQBE}
\end{figure}

For small condensate fractions (of less than about 10\%) the full dynamical
solution agrees well with the rate equation solution. However for larger
fractions, as illustrated in Fig.~\ref{Fig.MQBE}(a) 
we find that the bath dynamics have an appreciable effect. For this case,
where parameters are chosen to be appropriate for the MIT experiment and the
final condensate fraction is 24\%, the full dynamical solution (solid line)
grows faster than the rate equation solution (dotted line). This can be most
easily understood by introducing the concept of an effective chemical
potential $\mu _{\mathrm{eff}}$ for the thermal cloud (obtained by fitting a
Bose-Einstein distribution to the distribution function above $E_{R})$ and
using this in the simple growth equation Eq.(\ref{Eq.growth_n0}). Figure \ref%
{Fig.MQBE}(b) shows the evolution of both $\mu _{\mathrm{eff}}$ and $\mu
_{C}(n_{0})$ during the simulation, and we can see that during the
stimulation-dominated period of growth, $\mu _{\mathrm{eff}}$ substantially 
\textit{exceeds }the final equilibrium value, and thus from Eq.(\ref%
{Eq.growth_n0}), the backward rate from condensate to thermal cloud is
greatly inhibited. Ultimately this is due to the very severe cut that is
required on the initial equilibrium distribution in order to allow
evaporative cooling to proceed to such a high condensate fraction. We note
that Zaremba {\it et al.}\cite{Zaremba} have treated the full dynamical problem
using a somewhat different formalism, but obtain essentially the same
numerical results.

The theory reviewed in this section can be compared to the
experimental results of the MIT group\cite{MITgrowth}. It is found \cite%
{newbosgro,QK6,QK7} that there is very good agreement in some parameter
regimes, but significant discrepancies in others, which may be attributable
to the difficulty of fully characterising the experimental parameters. A\
definitive comparison is handicapped by the unavailability of a wide range
of primary experimental data, and further experiments will be required to
comprehensively test the theoretical predictions.

\section{\protect Output Coupling}

An output coupler for an atom laser is a device which extracts atoms from
the laser mode while preserving their coherence. All current coupler
mechanisms are based on applying an electromagnetic field to change the
internal state of the atom from a trapped to an untrapped state\cite%
{Savage93}. The first demonstration of an atom laser, by the MIT group\cite%
{MITOC97}, used an RF pulse to transfer atoms into untrapped zeeman
substates, where they then fell freely under gravity. A refinement of this
technique, using a precisely controlled trap and cw-RF fields has been used
to produce a quasi-cw coherent output beam\cite{MunichOC99}. The use of
Raman coupled output fields offers some advantages, principally in providing
a momentum kick to the outgoing atom beam\cite{Moy97b}, but also in that the
optical fields can be focussed into a small volume, achieving a measure of
spatial selectivity for the coupling. Recently the NIST group\cite{Hagley99}
demonstrated Raman output coupling in an experiment where the output beam
was ejected in the horizontal direction.

A number of mean field theoretical treatments have been given of RF\cite%
{Ballagh97,SteckOC98} and Raman couplers\cite{EdwardsOC99}, and elsewhere%
\cite{DurhamOC98} the effects of gravity have been included. Recently the
Oxford group\cite{JaphaOC99,ChoiOC99}, have given a detailed treatment of
output coupling from a finite temperature condensate. In this section we
review the theory of output couplers, beginning with a mean field treatment,
which establishes a framework for both RF and Raman output couplers. We then
briefly summarize the theory of finite temperature couplers.

\subsection{Mean field treatment of output coupling}

The basis of the mean field treatment of an output coupler is a pair of
coupled GP equations for the mean fields $\psi _{1}$ and $\psi _{2}$ of
atoms in internal states $|1\rangle $ and $|2\rangle $, as discussed
previously in section \ref{subsec.MF-models}. For convenience, we present a
dimensionless form of the equations 
\begin{eqnarray}
\frac{\partial \psi _{1}}{\partial t}\! &=&\!i\nabla ^{2}\psi
_{1}\!-\!iV_{1}\psi _{1}\!+\!i\frac{\lambda ^{\ast }(\mathbf{r},t)}{2}\psi
_{2}\!-\!iC\!\left[ |\psi _{1}|^{2}\!+\!w|\psi _{2}|^{2}\right] \!\psi
_{1}\!-\!iGy\psi _{1} , \label{Eq.GPOC} \\
\frac{\partial \psi _{2}}{\partial t}\! &=&\!i\nabla ^{2}\psi
_{2}\!-\!iV_{2}\psi _{2}\!+\!i\frac{\lambda (\mathbf{r},t)}{2}\psi
_{1}\!\!-\!iC\!\left[ |\psi _{2}|^{2}\!+\!w|\psi _{1}|^{2}\right] \!\psi
_{2}\!-\!iGy\psi _{2}\;.  \label{Eq.GPOCb}
\end{eqnarray}%
We shall assume that state $|1\rangle $ is the trapped state, with a
harmonic potential $V_{1}$ while state $|2\rangle $ is untrapped ($V_{2}=0$)
or anti-trapped ($V_{2}=-V_{1}$), so that $\psi _{2}$ represents the output
field. In Eqs.(\ref{Eq.GPOC}) and (\ref{Eq.GPOCb}) the units of time ($%
\tilde{t}_{0}$) and distance ($\tilde{x}_{0}$) are respectively $\omega
^{-1} $ and $(\hbar /2m\omega )^{1/2}$ where $\omega $ is the trap
frequency, and the fields are normalised such that $\int \!\left[ |\psi
_{1}|^{2}\!+\!|\psi _{2}|^{2}\right] d^{3}r=1$. The nonlinearity constant $C$
is $4\pi \hbar aN/(m\omega \tilde{x}_{0}^{3})$ (in 3D), and the relative
scattering length $w$ of the inter- to intra-species collisions is typically
close to 1. Gravity acts in the $-y$ direction, and the scaled acceleration
is $G=mg\tilde{x}_{0}/\hbar \omega .$ The coupling term $\lambda $ can
describe either a single photon coupling or a two photon (Raman) coupling,
and can be written in terms of a slowly varying envelope $\Omega (\mathbf{r}%
,t)$
\begin{equation}
\lambda (\mathbf{r},t)=\Omega (\mathbf{r},t)\exp i(\mathbf{k}_{em}\cdot 
\mathbf{r-}\Delta _{em}t)  \label{Eq.Omega_couple}
\end{equation}
where $\mathbf{k}_{em}$ and $\Delta _{em}$ measure the net momentum and net
energy transfer from the EM field to the output atom (in units $\hbar /%
\tilde{x}_{0}$ and $\hbar \omega $ respectively). For the case of a single
photon transition, $\Delta _{em}=\omega _{1}-\omega _{21},$ where $\omega
_{1}$ is the photon frequency, $\omega _{21}$ is the energy difference $%
E_{2}-E_{1}$ between states $|2\rangle $ and $|1\rangle $ (in the absence of
the traps), and $\Omega (\mathbf{r},t)$ is the Rabi frequency. For a Raman
transition connecting $|1\rangle $ to $|2\rangle $ (through a far detuned
intermediate state) via the absorption of a photon $\omega _{1}$ and the
emission of photon $\omega _{2}$ then $\Delta _{em}=\omega _{1}-\omega
_{2}-\omega _{21},$ and $\Omega (\mathbf{r},t)$ is equal to $\Omega
_{2}^{\ast }\Omega _{1}/\Delta $ where $\Omega _{j}$ is the Rabi frequency
associated with the $\omega _{j}$ radiation field, and $\Delta $ is the
detuning from the intermediate state. Our main interest here is in cw
coupling, and so $\Omega (\mathbf{r},t)$ will be independent of time. The
key difference between RF coupling and Raman coupling is that for RF
coupling, the momentum $\mathbf{k}_{em}$ is negligible and can be ignored.
Typically RF coupling (which may extend into the microwave region) is single
photon, but Raman coupling necessarily involves two photons. We note that in
Eqs.(\ref{Eq.GPOC}) and (\ref{Eq.GPOCb}) we have neglected the light shifts
that would occur for the two photon case, under the assumption that they
would be small compared to the potentials $V_{j}$ or the collisional mean
fields near the trap.

\begin{figure}[tbp]
\centering  \includegraphics[width=\textwidth]{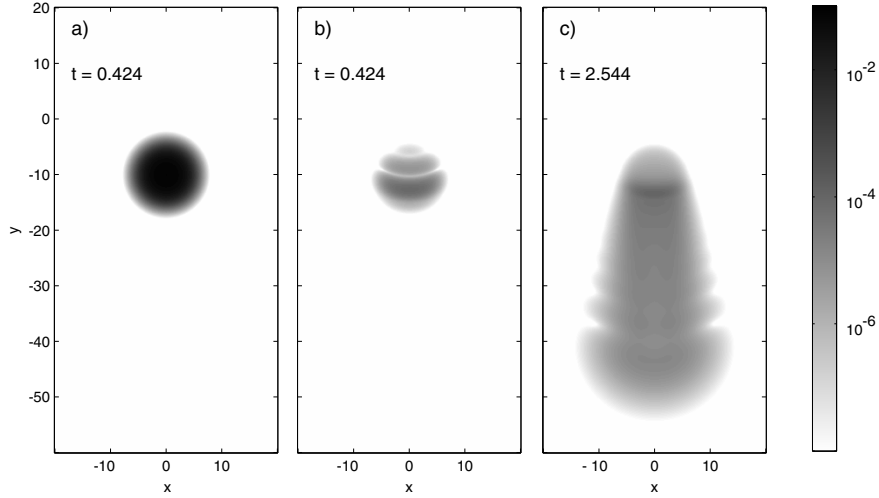}
\caption{{}Population densities in states $|1\rangle $ and $|2\rangle $
under cw output coupling in a gravitational field. (a) state $|1\rangle $ at 
$t=0.424$; (b) state $|2\rangle $ at $t=0.424;$ (c) state $|2\rangle $ at $%
t=2.544.$ Densities are plotted logarithmically. RF field is plane wave and
parameters are $\Omega =0.2,$ $\Delta _{em}=40,$ $G=5,$ $C=200,V_{2}=0.$}
\label{Fig.GPgravity}
\end{figure}

The coupler equations (\ref{Eq.GPOC}) and (\ref{Eq.GPOCb}) can be solved
numerically, and we illustrate in Fig.~\ref{Fig.GPgravity} a case for an RF\
coupler in two dimensions, where the system begins in an eigenstate of the
trap $V_{1}$, the RF field is plane wave, and $V_{2}=0.$ 
The first two frames show $|\psi _{1}|^{2}$ and $|\psi _{2}|^{2}$ soon after
the coupler has been turned on, and we note that the trapped state is
essentially unchanged from its initial state, while the output state (more
or less) copies the initial state. After some time, the output beam streams
downwards under gravity in a well directed uniform beam that resembles the
experimental results of the Munich group\cite{MunichOC99}.

\newpage
\subsection{Analytic treatments}

Many of the features of the output coupler behaviour that are found in
simulations of Eqs.(\ref{Eq.GPOC}) and (\ref{Eq.GPOCb}) can be understood in
terms of analytic approximations. The key parameters are the strength of the
coupling field ($\Omega $), the net energy match with the output states, and
the bandwidth of the coupling. When $\Omega $ is large (more precisely\cite%
{Ballagh97} $\Omega ^{2}+\Delta ^{2}\gg \mu (\mu +2|\Delta |)$, where $\mu $
is the chemical potential of the initial trapped state), it dominates the
remaining terms in Eqs.(\ref{Eq.GPOC}) and (\ref{Eq.GPOCb}), and the
resulting behaviour is synchronous Rabi cycling where each part of the
condensate oscillates between being fully in state $|1\rangle $ and fully in
state $|2\rangle $, at frequency $\Omega .$ The atoms spend such short
periods of time in the untrapped state, that they are unable to escape. On
the other hand when $\Omega $ is small compared to $\mu $, other terms in
Eqs.(\ref{Eq.GPOC}) and (\ref{Eq.GPOCb}) become more significant, and the
difference between the trap potentials becomes important. We can understand
this by transforming Eqs.(\ref{Eq.GPOC}) and (\ref{Eq.GPOCb}) into a rotating
frame using $\tilde{\psi}_{2}=\psi _{2}\exp (i\Delta _{em}t),$ neglecting
kinetic energy (because $t$ is small and the atoms do not have time to
move), and then recognising that the equations simply become Rabi cycling
with a spatially dependent effective detuning%
\begin{equation}
\Delta _{\mathrm{eff}}(\mathbf{r)}=\Delta _{em}-(V_{1}-V_{2})\,.
\label{Eq.Delta_eff}
\end{equation}%
The mean field contribution to $\Delta _{\mathrm{eff}}$ cancels (for $w=1$)
because the atom feels the same mean field in either state $|1\rangle $ or $%
|2\rangle .$ Now, because the spectral width $\Omega $ of the Rabi cycling
is small, significant transfer from $|1\rangle $ to $|2\rangle $ will only
occur in a small spatial region near the value of $\mathbf{r}$ for which $%
\Delta _{\mathrm{eff}}(\mathbf{r)}=0.$ The resulting spatially selective
output coupling can be seen in Fig. \ref{Fig.GPgravity} (b) and (c), where
the preferred region of coupling is a narrow crescent near $y=-10.$ The
location of this region can be calculated by noting that the magnetic traps
(i.e. $V_{1}$ and $V_{2}$) are centered on $x=0,y=0,$ and thus the condition 
$\Delta _{\mathrm{eff}}(\mathbf{r)}=0$ for a given choice of $\Delta _{em}$
is equivalent to selecting an equipotential line for the potential
difference $V_{1}-V_{2},$ which is a circle centered on $x=0,y=0.$ Due to
gravity, the equilibrium centre of mass position of condensate $\psi _{1}$
drops below the centre of the magnetic trap to a position where the spring
force is equal to the gravitational force, as shown in Fig. \ref%
{Fig.GPgravity}(a). Thus the spatially selected region of coupling, where
the trap equipotential intersects the trapped condensate, takes a crescent
shape. Of course the coupling time $\tau _{c}$ must be long enough that
transient broadening does not obscure the spatial selectivity.

\subsection{Weak coupling regime}

The weak coupling regime is of most interest for cw atom lasers, and
perturbative solutions have been derived which describe key aspects of the
output field behaviour\cite{EdwardsOC99,JaphaOC99}. Following those
treatments we make the assumption of an undepleted trapped state and a very
dilute output state, which allows us to replace $\psi _{1}$ in Eq.(\ref%
{Eq.GPOCb}) by its initial value $\psi _{1}(t=0)$ and to neglect the
meanfield interaction of the output state with itself. Taking a Thomas-Fermi
approximation for $\psi _{1}(t=0)$, Eq.(\ref{Eq.GPOCb}) becomes a linear
Schrodinger equation 
\begin{equation}
\frac{\partial \psi _{2}}{\partial t}\!=-iH_{oc}\psi _{2}+\!i\frac{\lambda (%
\mathbf{r},t)}{2}\Psi _{TF}\exp (-i\mu t)\!\! , \label{Eq.linOC}
\end{equation}%
where $H_{oc}=-\nabla ^{2}+V_{oc}$\thinspace . The potential for the
outcoupled wave is $V_{oc}=V_{2}+w(\mu -V_{1})+Gy$ and the final term in Eq.(%
\ref{Eq.linOC}) is a source term in which the Thomas-Fermi wavefunction is $%
\Psi _{TF}=[\mu -V_{1}]^{1/2}.$ The eigenfunctions $\phi _{E}(\mathbf{r})$
of $H_{oc}$ provide a convenient basis for solving Eq.(\ref{Eq.linOC}).
Taking for simplicity the one dimensional case (and neglecting gravity) we
obtain the solution
\begin{equation}
\psi _{2}(x,t)=i\int dE\left\{ e^{i(B-E)t}
\left[ \frac{\sin Bt}{B} \right] \right\}
\Lambda (E)\phi _{E}(x)\,,  \label{Eq.OCweak_explicit}
\end{equation}
where $B=[E-(\mu +\Delta _{em})]/2$ and 
\begin{equation}
\Lambda (E)=\frac{1}{2}\int dx\phi _{E}^{\ast }(x)\Omega (x)\exp
(ik_{em}x)\Psi _{TF}\,,  \label{Eq.OCoverlap}
\end{equation}%
is the matrix element of the EM coupling amplitude between the output
eigenfunction and the Thomas-Fermi trap state. Eq.(\ref{Eq.OCweak_explicit})
provides useful insight into the properties of the output field. The term in
square brackets is a familiar factor from perturbation theory, and at small
times the term in braces $\left\{ {}\right\} \longrightarrow t$ so that $%
\psi _{2}(x,t)\approx it[\Omega (x)/2]\exp (ik_{em}x)\Psi _{TF}(x).$ Thus at
early times the weak coupler produces an output field that is a copy of the
initial trapped state, spatially masked by the EM field amplitude. At long
times the term in square brackets $[\,]\longrightarrow 2\pi \delta (E-\mu
-\Delta _{em})$ so that the output state becomes proportional to the
eigenfunction $\phi _{E}(x)$ 
\begin{equation}
\psi _{2}(x,t)\approx i2\pi \exp (-iE)t\Lambda (E)\phi _{E}(x)
\end{equation}%
where $E$ conserves energy according to 
\begin{equation}
E=\mu +\Delta _{em}\,,\,  \label{Eq.Energy_output}
\end{equation}%
and the function $\Lambda (E)$ plays the role of a spectral filter. The low
energy cutoff of $\Lambda (E)$ can determined\cite{EdwardsOC99} by
consideration of the classical turning points of the wavefunction $\phi
_{E}(x),$ which shows that $\Lambda (E)=0$ for $E\leq 0$. Thus (for the case 
$V_{2}=0$), the lowest possible output energy for an atom is $E=0$, which
gives a corresponding lower limit for the coupling field detuning $\Delta
_{em}=-\mu ,$ [see Eq.(\ref{Eq.Energy_output})] at which value the EM field
absorbs all the energy $\mu $ of the trapped condensate atom. It is more
difficult to find the upper cut-off of $\Lambda (E),$ but the spatial
selectivity criterion given by Eq.(\ref{Eq.Delta_eff}) suggests that output
coupling will cease when $\Delta _{em}$ exceeds the largest value of $%
V_{1}-V_{2}$ in the condensate. Setting $V_{2}=0,$ neglecting gravity, and
using the Thomas-Fermi radius for the edge of the condensate, we can then
write the limits for which $\Delta _{em}$ produces effective coupling as 
\begin{equation}
-\mu \leq \Delta _{em}\leq \mu \,.  \label{Eq.DeltaRange}
\end{equation}%
The bandwidth of the output coupling $\Delta \omega _{B}$, which is
determined primarily by $\Lambda (E),$ determines the time scale for the
output field to reach its steady state ($\sim \lbrack \Delta \omega
_{B}]^{-1}$). For the case of a Raman coupler (for which $\exp (ik_{em}x)$
must be retained in Eq.(\ref{Eq.OCoverlap})) or a gravitational field, $%
\Delta \omega _{B}$ is appreciably larger than for the case of a simple RF
coupler in free space. When $\Omega \ll \Delta \omega _{B}$, and thereby
Rabi cycling is suppressed, the faster untrapped atoms escape%
\cite{ChoiOC99}.

\subsection{Finite Temperature Output coupling}

The Oxford group have extended the theory of output coupling to include the
description of finite temperature 
condensates\cite{JaphaOC99,ChoiOC99} (see the article by Burnett in 
this volume).
Their treatment, which uses operator fields with a number conserving
approach is sophisticated, but has an underlying structure that is very
similar to that presented in the previous section. The atom has two internal
states $|t\rangle $ and $|f\rangle $ (trapped and free) and the atom field
operator is divided into $\hat{\psi}_{t}(\mathbf{r},t)$ for the trapped
state and $\hat{\psi}_{f}(\mathbf{r},t)$ for the output field,
(corresponding to $\psi _{1}$ and $\psi _{2})$. The same initial
approximations are made in assuming the output coupling is weak so that the
trapped field is undepleted and the output field is very dilute. Inelastic
collisions are neglected, and an EM coupling field with amplitude given by
Eq.(\ref{Eq.Omega_couple}) models either RF or Raman coupling. This allows
an equation to be written for $\partial \hat{\psi}_{f}(\mathbf{r})/\partial
t $, which is completely analogous to Eq.(\ref{Eq.linOC}). The output field
moves in a potential comprised of the trap potential for state $|f\rangle $
and the meanfield of the (unmodified) initial trapped state, while the self
field of $\hat{\psi}_{f}$ is neglected. The source term involves the
equilibrium field operator for the trapped condensate, $\hat{\psi}_{t}(\mathbf{r,}%
t=0)$ but it is important to note that this is not a meanfield, but includes
a thermal component, in which the quasiparticle levels are assumed to be
occupied according to a Bose-Einstein distribution of temperature $T$ and
chemical potential $\mu $. The operator $\hat{\psi}_{f}$ is expanded in
modes $\phi _{\mathbf{k}}(\mathbf{r})$ of energy $E_{\mathbf{k}}$ that are
eigenfunctions of a Hamiltonian that is equivalent to $H_{oc},$ and leads to
the solution for the free field 
\begin{eqnarray}
\hat{\psi}_{f}(\mathbf{r},t) &\approx &-i\Psi _{f}^{0}(\mathbf{r},t)\hat{a}%
_{0}(0)  \nonumber \\
&&-\hat{a}_{0}(0)\frac{i}{\sqrt{\hat{N}_{0}}}\sum_{j}\left[ \Psi _{f}^{j+}(%
\mathbf{r},t)\hat{\alpha}_{j}(0)+\Psi _{f}^{j-}(\mathbf{r},t)\hat{\alpha}%
_{j}^{\dag }(0)\right] ,  \label{Eq.Psi_f_op}
\end{eqnarray}%
where $\hat{a}_{0}$ is the annihilation operator for the condensate mode, $%
\hat{N}_{0}$ is the condensate occupation operator $\hat{a}_{0}^{\dagger }%
\hat{a}_{0},$ and $\hat{\alpha}_{j}$ is the annihilation operator for the $%
j^{th}$ quasiparticle level, of energy $\mu +E_{j}.$ The mode functions in
Eq.(\ref{Eq.Psi_f_op}) are 
\begin{equation}
\Psi _{f}^{\eta }(\mathbf{r},t)=\sum_{\mathbf{k}}\varphi _{\mathbf{k}}(%
\mathbf{r})D_{\mathbf{k}\eta }(t)\Lambda _{\mathbf{k}\eta }e^{-iE_{\mathbf{k}%
}t}  \label{Eq.Psi_f_mode}
\end{equation}%
for $\eta =0,\,j+,\,j-$\thinspace . The quantity $D_{\mathbf{k}\eta
}(t)=i(e^{-i(E_{out}^{\eta }-E_{\mathbf{k}})t}-1)/(E_{out}^{\eta }-E_{%
\mathbf{k}})$ is an energy selective function in which 
\begin{equation}
E_{out}^{\eta }=\mu +\Delta _{em}+E_{\eta }\,,  \label{Eq.finite_T_energy}
\end{equation}%
with $E_{j\pm }=\pm E_{j}$ , and $E_{0}=0.$ The quantity $\Lambda _{\mathbf{k%
}\eta }$ is a frequency filter analogous to $\Lambda (E)$ in Eq.(\ref%
{Eq.OCoverlap}) and is given by

\begin{equation}
\Lambda _{\mathbf{k}\eta }=\frac{1}{2}\int d^{3}\mathbf{r}\varphi _{\mathbf{k%
}}^{\ast }(\mathbf{r})\Omega (x)\exp (ik_{em}x)\xi _{\eta }(\mathbf{r})
\end{equation}%
where $\xi _{\eta }$ for $\eta =0,\,j+,\,j-$, represents $\psi _{0},$ $u_{j}$%
, $v_{j}^{\ast }$, the trapped condensate wavefunction, and the usual
quasiparticle amplitudes, respectively. We can now interpret the three terms
in Eq.(\ref{Eq.Psi_f_op}). The first term is the condensate output and
corresponds to the weak coupling mean field calculation given by Eq.(\ref%
{Eq.OCweak_explicit}); the second term ($\Psi ^{j+}$) represents an atom
ejected from a trapped quasiparticle mode (stimulated quantum evaporation of
the thermal excitations) and the third term ($\Psi ^{j-}$) represents an
atom ejected from the condensate and creating a trapped quasiparticle
(pair-breaking). Each of these processes obeys energy conservation given by
Eq.(\ref{Eq.finite_T_energy}), which for the condensate term ($\eta =0$) is
identical to Eq.(\ref{Eq.Energy_output}). For the stimulated evaporation of
a quasiparticle ($\eta =j+$) we see that the output atom has energy equal to
that of the initial trapped quasiparticle plus the energy provided by the EM
field. For the pair breaking term, an atom comes out of the condensate, but
leaves behind an extra quasiparticle, so that the output energy is $\mu
-E_{j}$ plus the energy provided by the EM field.

\begin{figure}[tbp]
\centering  \includegraphics[width=6cm]{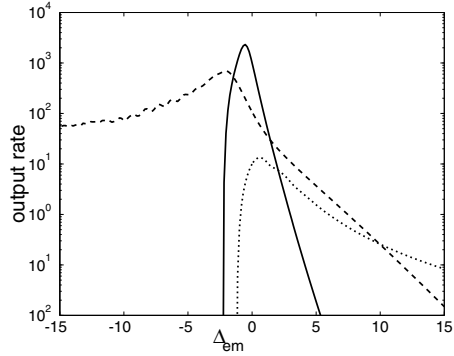}
\caption{Population output rate from a finite temperature condensate with an
RF coupler. Solid line, condensate component. Dashed line, stimulated
quantum evaporation component. Dotted line, pair breaking component.
Parameters: $T=0.5T_{c}$ $(150\hbar \protect\omega /k)$, $\protect\mu %
=2.3\hbar \protect\omega /k$ , $N_{t}=2000$ atoms.}
\label{Fig.ChoiFig}
\end{figure}

The density of output atoms is given by 
\begin{equation}
n_{out}(\mathbf{r},t)=\langle \hat{\psi}_{f}^{\dag }(\mathbf{r},t)\hat{\psi}%
_{f}(\mathbf{r},t)\rangle  \label{Eq.nout}
\end{equation}%
and by assuming that all cross correlations between different operators $%
\hat{a}_{0},\hat{\alpha}_{j},\hat{\alpha}_{j}^{\dag }$ vanish and only the
populations $\langle \hat{a}_{0}^{\dag }\hat{a}_{0}\rangle =n_{0},$ $\langle 
\hat{\alpha}_{j}^{\dag }\hat{\alpha}_{j}\rangle =n_{j},$ and 
$\langle \hat{\alpha}_{j} \hat{\alpha}_{j}^{\dag} \rangle =n_{j}+1$ remain, 
we can use Eq.(\ref{Eq.Psi_f_op}) to
write the output density

\begin{equation}
n_{\mathrm{out}}(\mathbf{r},t)=n_{0}|\Psi _{f}^{0}(\mathbf{r}%
,t)|^{2}+\sum_{j}\left[ n_{j}|\Psi _{f}^{j+}(\mathbf{r},t)|^{2}+(n_{j}+1)|%
\Psi _{f}^{j-}(\mathbf{r},t)|^{2}\right] ,
\end{equation}%
where the three separate components of the output field are easily
identified. The time evolution of the output field is governed by the output
bandwidth $\Delta \omega _{B}$ as discussed in the previous section, and
first discussed extensively by Choi {\it et al.}\cite{ChoiOC99}. In the weak
coupling limit, $\Omega \ll \Delta \omega _{B},$ we find once again that for
short times the output state copies the trapped state $\xi _{\eta }(\mathbf{r%
})$ 
\begin{equation}
\Psi _{f}^{\eta }(\mathbf{r},t)\sim _{t\rightarrow 0}[\Omega (\mathbf{r}%
)/2]\xi _{\eta }(\mathbf{r})t\,.  \label{Eq.Psi-short}
\end{equation}%
For long times, energy conservation selects only output states for which $E_{%
\mathbf{k}}=E_{out}^{\eta }.$ Choi {\it et al.}\cite{ChoiOC99} have given a
calculation for RF output coupling in one dimension, from a condensate at 
temperature $T=0.5T_{c}$. They calculate the rate $dN_{\mathrm{out}}/dt$ in
steady state, where 
\begin{equation}
N_{\mathrm{out}}=\int n_{\mathrm{out}}(\mathbf{r},t)d^{3}r
\end{equation}%
and the result is presented in Fig.~\ref{Fig.ChoiFig}, where the three
components of the output are plotted (as different line types) as a function
of the EM detuning $\Delta _{em}$. 
A\ key feature of the output coupling that is readily apparent in this
figure is how the detuning $\Delta _{em}$ controls the character of the
output field. For $\Delta _{em}< -\mu ,$ only the thermal quasiparticles are
ejected from the trap, as we expect from energy conservation (see Eq.(\ref%
{Eq.finite_T_energy})): only the quasiparticles have sufficient energy to
absorb this amount of energy from the radiation field. For $\Delta
_{em}>-\mu ,$ up to $\Delta _{em} \approx \mu ,$ fully coherent
condensate dominates the output (see also Eq.(\ref{Eq.DeltaRange})). Finally
for $\Delta _{em} >\approx -\mu ,$ both pair breaking (where the energy of
the initial condensate atom plus the photon can be left in the trap) and
thermal evaporation can occur.

As a final comment, we note that the processes of stimulated thermal
evaporation and pair breaking reduce the coherence of the output field. Choi
{\it et al.}\cite{ChoiOC99} have calculated the coherence functions $%
g^{(1)}(x_{1},x_{2},t)$ and $g^{(2)}(x_{1},x_{1},t)$ and shown that even in
the condensate dominated regime, the first order coherence is degraded, and $%
g^{(2)}$ exhibits symptoms of atom bunching.

\section{Acknowledgements}

RJB is grateful for many instructive conversations
over the past few years with Professors C. Gardiner and K. Burnett, which
have helped shape this article. This research was supported by the Marsden
Fund of New Zealand under contract PVT902.

CMS acknowledges collaborations with J. Hope, G.Moy, and N. Robins. 
Our work is supported by the Australian Research Council.

%%%%%%%%%%%%%%%%%%%%%%%%%%%%%%%%%%%%%%%%%%%%%%%%%%%%%%%%%%%%%%%%%%%%%%%

\end{document}